\documentclass{aa}  

\usepackage{graphicx}
\usepackage{txfonts}
\usepackage{hyperref}
\usepackage{subfigure}
\usepackage[dvipsnames]{xcolor}

\begin{document}

   \title{Mass transport in a moist planetary climate model}

   %\subtitle{Moist Dynamical Core}

   \author{J. M. Mendon\c ca
          \inst{1}\thanks{joao.mendonca@space.dtu.dk\newline   Homepage: software-oasis.com}
          }

   \institute{$^{1}$ National Space Institute, Technical University of Denmark, Elektrovej, 2800, Kgs. Lyngby, Denmark
                }
 
  \abstract
   {
Planetary climate models (PCMs) are developed to explore planetary climates other than that of the Earth. Therefore, the methods implemented need to be suitable for a large diversity of conditions. Every planet with a significant atmosphere has condensible cycles (e.g. the hydrological cycle), which can play an essential role in the planet's appearance and environment. We must accurately represent a condensible cycle in our planet simulations to build a powerful planetary climate predictor. \texttt{OASIS} is a 3D PCM capable of self-consistently representing the main physical processes that drive a planet's environment. In this work, we improve the representation of mass transport in \texttt{OASIS}, which is the first step towards a complete and flexible implementation of a condensible cycle. We implement an upwind-biased scheme on a piece-wise linear approximation with a flux limiter to solve the mass transport equation. We first benchmark the new scheme on a 2D problem that confirms the superior properties of the new method over the central finite-volume method in terms of performance, accuracy, and shape-preserving mass distribution. Due to the new scheme's less dispersive nature, we do not have to apply any unphysical diffusion to maintain the model stable. \texttt{OASIS} includes the new improved solver in the total mass and the tracer (e.g. clouds and individual gas chemical species) transport. We couple the new formulation with physical schemes and validate the new code on two 3D simulations of an ocean Earth-like planet and an ocean tidally locked planet. The new \texttt{OASIS} simulations are robust and do not show any known problems from the dynamics-physics coupling. We show that the two simulations capture the main characteristics of ocean planet atmospheres and are easy to set up. We propose these two simulations as the first standard benchmark tests for models built to explore moist planetary environments.
   }


   \maketitle
%
%-------------------------------------------------------------------

\section{Introduction}
Planetary climate models (PCMs) are flexible numerical tools capable of representing vast planet conditions. Every detail on how equations are solved in the models has repercussions on the model output and modelled physics. In this work, we focus on the atmospheric mass transport problem and simple moist processes. 

A gas can exist in multiple phases in a planet environment.  It can, for example, condense and form clouds or rain and fall onto the planet's surface.  In this work, we define the general term condensible cycle as the `life' cycle of a condensible gas in a planet's environment. Any planet with a significant atmosphere is expected to host a least one condensible gas. The composition and spatial distribution of the condensible material are tightly connected with the context of the planetary environment. To have a deep understanding of how a climate of a particular planet works, we need to study the impact of all the possible condensible cycles for the specific planet. On Earth, it is the water cycle that plays a vital role in the planet's climate.  The condensible gases are expected to be diverse under different planet conditions, for example water on Earth, sulphuric acid on Venus (e.g. \citealt{2011Taylor}), and methane on Titan (e.g. \citealt{2017Horst}). By having a robust representation of moist processes in the atmosphere, we improve our model predicting power in planetary climates. In this work we present the first steps towards a complete implementation of a condensible cycle in the \texttt{OASIS} model (\citealt{2020Mendonca}).

Atmospheric circulation transports gas and clouds across the atmosphere. A robust and fast representation of the atmospheric transport in 3D simulations is essential to avoid compromising the physics in the simulations. Inadequate atmospheric transport models can result in a loss of mass conservation, the creation of artificial high-frequency waves, or a poor representation of how chemistry and clouds are redistributed across the atmosphere. Condensible constituents of the atmosphere interact in various ways with the radiation. The distribution of the components can have a substantial impact on the atmosphere's energy budget (e.g. the planet's bond albedo; \citealt{2016Read}). The heat absorbed or released during the different phase changes also impacts atmospheric thermodynamics and has an important role in the planet's atmospheric circulation. 

Some phenomena that involve complex representations of moist processes remain poorly explored by 3D simulations, such as atmospheric collapse (e.g. \citealt{2015Wordsworth}) and runaway greenhouse phenomena (e.g. \citealt{1969Ingersoll}). These phenomena require the representation of a complete condensible cycle, such as cloud microphysics and condensate transport, which can be the main component of the atmosphere. The thermodynamical equations in the model's dynamical core need to capture the leading physics and avoid including assumptions such as those typically adopted from Earth climate models to facilitate the numerical schemes. One such assumption is the limit of small condensable content in the atmosphere compared to the main atmospheric component (\citealt{2017Ullrich}), such as water and nitrogen on Earth. The limit of the small condensable abundance simplifies the vertical integration of the equations in the dynamical core of Earth models since quantities such as gas constant and heat capacity remain fixed. 

Clouds or hazes in terrestrial planets can mask the atmosphere below and challenge the detection of molecules in the atmosphere (e.g. \cite{2019Fauchez}, \cite{2020Suissa}, \cite{2020Komacek}), which highlights the need to have a robust aerosol distribution in 3D numerical simulations to maximise the science output of observational campaigns. We need to have a better understanding of the limitations of our models. Poor representations of these atmospheric phenomena in the simulations can lead to a flawed interpretation of the observational data and analysis of unphysical phenomena from the 3D simulations. The details on how the clouds and chemistry are being transported in 3D models of planetary simulations are often left out of scientific papers, making it hard to compare our model results with other models applied to extra-solar planets. Every formulation has pros and cons, and 3D planetary climate users or developers should make more explicit the options adopted in their simulations. This work aims to give the first steps towards developing a complete moisture cycle in a 3D dynamical core, focusing on the problem of mass transport and latent heat release. We explain how the modeller can estimate the numerical schemes' errors and identify the primary source of numerical errors, which is essential for future model inter-comparison studies. By solving the total mass and tracer equations with the same solver, we maintain consistency between the atmospheric mass and any tracer in the atmosphere in terms of spatial distribution and accuracy, which becomes important for planetary conditions where the condensable gas has a large concentration in the atmosphere. As far as we know, in the planetary community only the model developed in \cite{2020Ge} can follow a similar approach due to the higher-order scheme implemented to solve the mass conservation equation.

In the next section we describe the new equations and numerical schemes implemented in the \texttt{OASIS} dynamical core (known as \texttt{THOR}); we also explore the accuracy of the new advection scheme.  In Sect. \ref{sec:3dbenchmark_test} we present the simplified physics implemented in the 3D simulations, and in Sect. \ref{sec:results_3d} we present the results of the 3D ocean planet benchmark tests. Concluding remarks are presented in Sect. \ref{sec:concl}.

\section{New equations in the \texttt{OASIS} dynamical core}
\label{sec:new_oasis}
\texttt{OASIS} is a 3D virtual-lab to study planetary environments that has been written from ground-up (\citealt{2020Mendonca}). This platform comprises different physics and chemistry modules that interact with each other to reproduce self-consistent results. \texttt{OASIS} makes use of the flexible dynamical core, \texttt{THOR}, which has a 3D non-hydrostatic nature aimed at exploring a large diversity of planets (\citealt{2016Mendoncab} and \citealt{2020Deitrick}). One example of the \texttt{OASIS} robustness and success is the recent successful \texttt{OASIS} simulations of Venus (\citealt{2020Mendonca}). The climate of Venus is very challenging to simulate due to its highly reflective clouds that entirely cover the planet and its massive atmosphere that requires tens of thousands of simulation days to reach thermal equilibrium (e.g. \citealt{2010Lebonnois} and \citealt{2016Mendoncaa}). Venus has strong winds that cause the atmosphere to rotate 60 times faster than the solid planet, in a phenomenon called `super-rotation' (e.g. \citealt{1986Read}, \citealt{2017Lavegaa}). The strong winds have an enormous impact on the planet's climate. Using our state-of-the-art PCM, \texttt{OASIS}, we were able to simulate the extreme conditions observed in the Venus cloud region (\citealt{2020Mendonca}).

In this work, we mainly expand the scope of the atmospheric dynamics module (\texttt{THOR}) to represent the atmospheric transport in a more accurate and physically based way to couple the dynamical core with simple physical representations of the atmospheric water cycle. 

\subsection{Expanding \texttt{THOR} equations}
\texttt{THOR} is the module that solves the resolved atmospheric fluid flow (\citealt{2016Mendoncab}). \texttt{THOR} was built to be flexible and efficient in exploring planetary atmospheres, and the equations solved for dry atmospheres are presented in \cite{2016Mendoncab} and \cite{2020Deitrick}. The equations of motion are solved in a modified icosahedral grid (\citealt{2001Tomita}). In this work, we extend and modify the code \texttt{THOR} to include:
(i) a tracer transport equation explicitly in the main set of equations of the dynamical core and (ii) improved consistency and accuracy of the mass transport representation.

The new set of equations solved in the dynamical core are the following:
\begin{align}
    &\frac{\partial}{\partial t}(\rho) + \nabla\cdot(\rho \textbf{v}) = 0,\label{eq:thor1}\\
    &\frac{\partial \rho\textbf{v}}{\partial}+\nabla\cdot(\rho\textbf{v}\otimes \textbf{v}) = -\nabla P-\rho g \hat{\textbf{r}}-2\rho \textbf{$\Omega$}\times\textbf{v},\label{eq:thor2}\\
    &\frac{\partial}{\partial t}(\rho \theta) + \nabla\cdot(\rho  \theta \textbf{v}) = \frac{Q\rho\theta}{T c_p}, \label{eq:thor3}\\
    &\frac{\partial}{\partial t}(\rho q_i) + \nabla\cdot(\rho q_i \textbf{v}) = S_{q_i}, 
    \label{eq:thor4}
\end{align}
where $\rho$ is the atmospheric density, $\textbf{$v$}$ is the velocity, $P$ is the atmospheric pressure, $g$ is the planet gravity, \textbf{$\Omega$} is the planet's rotation vector, $\theta$ is the potential temperature, $T$ is the atmospheric temperature, $c_p$ is the heat capacity of the atmosphere, $Q$ is the heating or cooling from radiation, convection, condensation and evaporation, $q_i$ is the concentration of a tracer with index $i$ (e.g. clouds, chemical gases), and $S_q^i$ is the physical source or loss of the tracer $i$. In our new platform the prognostic variables are: $\rho \textbf{u}_h$, $\rho w$, $\rho$, $\rho \theta$, and $\rho q_i$. In addition to the equations above, we assume that the ideal gas equation and the dry atmosphere's thermodynamic equations are valid anywhere in the atmosphere. We have not further explored the impact of moist thermodynamic equations because we want to focus only on the impact of the atmospheric mass transport problem in this work. A complete coupling of the thermodynamic equations is planned for follow-up work. The atmosphere's specific heat capacities and gas constants vary for composition, phase changes, and temperature variations in a moist atmosphere. The equations or methods have to be adapted to the needed atmospheric conditions. Recent Venus simulations (e.g. \citealt{2010Lebonnois} and \citealt{2016Mendoncaa}) are good examples on how variations in the specific heat capacity as a function of temperature can be incorporated in 3D PCMs. In addition, planetary conditions where the main atmospheric gas is condensing will require more complete equations than those used in Earth climate models.

As explained in \cite{2016Mendoncab} and \cite{2020Deitrick}, \texttt{THOR} discretises the equations horizontally in an Arakawa A-grid and the vertical levels on a non-staggering grid. The spatial integration of the momentum and entropy equations are solved using the central finite-volumes methods from \cite{2002Satoh}, and \cite{2004Tomita}. In this work, we upgrade the spatial integration of the density and tracer equations with a new upwind-biased scheme with a flux limiter (\citealt{2007Miura}), as explained in the section below. The density and the tracers are solved with the same solver to maintain physical consistency.

The time integration is explained in \cite{2016Mendoncab}. Our numerical scheme involves splitting the time integration into short explicit steps for the terms associated with acoustic modes and large steps for the rest (e.g. \citealt{2002Wicker} and \cite{2008Skamarock}). The fast acoustic waves can compromise the stability of the model. The equations are also formulated to solve the vertical component of the momentum using an implicit method (HE-VI method - e.g. \citealt{2004Tomita}). The implicit method in the vertical direction is necessary due to the often much higher spatial resolution in the vertical direction compared to the horizontal spacing.

\subsection{Tracer transport}
\label{subsec:tracer_transp}
The tracer\footnote{In this work, a tracer is a quantity that follows the atmospheric flow, such as clouds or gas chemical species.} transport scheme is one of the most important routines in a climate model dynamical core. Equation \ref{eq:thor4} shown above represents the tracer transport. The right side term of the equation, $S_q$, is the physical source or sink of the tracer (e.g. condensation and evaporation). If $S_q$ is zero, the tracer concentration is conserved. There are many ways of solving the tracer transport equation, and in this work, we improved the existing tracer transport in the dynamical core \texttt{THOR} (\citealt{2016Mendoncab}). The original scheme implemented in \texttt{THOR} to represent the tracer transport is a central finite-volume scheme from \cite{2001Tomita}. This scheme is well known to be dominated by numerical dispersion that creates high-frequency patterns in the solutions (e.g. \citealt{1999Durran}). The non-physical solutions could create maxima and minima artificially and hence produce non-physical scenarios (e.g. rain produced by an overshoot in the cloud transport scheme). To overcome this numerical inaccuracy, we have in \cite{2018Mendoncab} advected chemical species in the atmosphere of hot Jupiter planets with a hyper-diffuse term in the tracer transport equation. The diffuse term removes the high-frequency oscillations created artificially by the central finite volume. The hyper-diffuse term in the tracer transport helps to maintain the simulation stable, but the formulation is not physical-based for mass transport. The new flux divergence operator implemented here was developed in \cite{2007Miura}. The new method is an upwind-biased advection scheme on a piece-wise linear approximation, which we also complement with a flux limiter developed in \cite{1996Thuburn}. These two methods allow us to achieve two essential properties in the tracer transport scheme: mass conservation (from the finite-volume scheme) and shape-preserving mass distribution (from the flux-limiter method). The flux-limiter method removes, for example, over- or under-shoot solutions, and is important to ensure the monotonicity of the gradient concentration. 

To solve Eq. \ref{eq:thor4}, we first apply the finite-volume discretisation from \cite{2007Miura}:
\begin{equation}
    \rho_0^{t+\Delta t}q_0^{t+\Delta t}=\rho_0^{t}q_0^{t}-\frac{\Delta t}{A_0}\sum_{i}^{N_s}(l_i\tilde{\rho}_{\textbf{R}_i}\tilde{ q}_{\textbf{R}_i}\textbf{v}_{\textbf{R}_i}^{t+\Delta t/2}\cdot\textbf{n}_i)
\label{eq:miura07}
,\end{equation}
where $\rho$ is the atmospheric density, $q$ the tracer concentration, $A_0$ is the control cell area, $\Delta t$ is the large time step in the dynamical core, $l$ is the length of the pentagon or hexagon sides, $\textbf{v}$ is the wind velocity, $N_s$ is the number of sides of the pentagons or hexagons, and $n$ is the normal vector at each side of the control volume. The indices $0$ in Equation \ref{eq:miura07} refers to quantities defined at the centre of the control volumes and $\textbf{R}$ at cell face centres. \cite{2007Miura} provides more details about the equations and discretisation. The upwind scheme alone does not maintain the monotonicity of the tracer distribution. To guarantee the monotonicity we apply the flux limiter from \cite{1996Thuburn}. The same procedure was used in \cite{2007Miura}. All the details on how to implement the flux limiter used in this work are explained in \cite{1996Thuburn}.

\begin{figure}[ht]
\includegraphics[width=1.0\columnwidth]{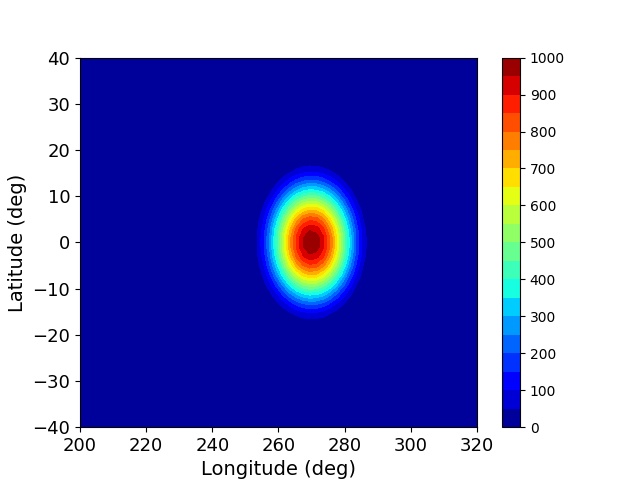}
\caption{Initial cosine bell shape defined in Eq. \ref{eq:cosinebell}. The colour map is in arbitrary units.} 
\label{fig:bell}
\end{figure}

To guarantee that the new atmospheric transport solver is well implemented in \texttt{THOR} we perform the test case 1 described in \cite{1992Williamson}. The idea of the test is simple, where a cosine bell shape distribution of a tracer is transported across the globe until its initial position. The simplicity of the test allows us to easily find mistakes in our implementation. The non-divergent wind that transports the tracer in our test is set to a solid body rotation,
\begin{align}
        u(\phi, \lambda) &= u_0\cos(\lambda)\\
        v(\phi, \lambda) &= 0,
\end{align}
where $u$ and $v$ are the zonal and meridional wind directions, respectively, $\phi$ is the longitude and $\lambda$ is the latitude. The variable $u_0$ is defined as:
\begin{equation}
    u_0=\frac{2\pi R_{planet}}{P_{bell}}.
    \label{eq:u0}
\end{equation}
The constant $R_{planet}$ is the radius of the planet and $P_{bell}$ the period that the cosine bell shape takes to do a complete revolution around the planet. We set these two constants to a radius of $6.371\times10^6$ m and 12 days period. The cosine bell function was constructed following \cite{1992Williamson} work:
\begin{equation}
    h(r) = \begin{cases}(h_0/2)[1+\cos{\pi r/R}] &\text{if $r < R$}\\
0 &\text{if $r \ge R$},
    \end{cases}
\label{eq:cosinebell}
\end{equation}
where $r$ is the great circle distance between any point and the centre of the cosine bell. The variable $h$ represents the concentration of an arbitrary tracer. The initial centre for the cosine bell is $(longitude, latitude) = (270^\circ, 0^\circ)$ . Figure \ref{fig:bell} shows a map of the initial cosine bell shape. Other constants in Eq. \ref{eq:cosinebell} were set to $h_0 = 1000$ m and $R = R_{planet}/3$. Other configurations of the non-divergent wind and initial position of the cosine bell distribution were explored with no significant impact in the final conclusions about the solvers accuracy.

\begin{table}
  \caption{Time step and spatial resolutions for the cosine bell test cases using different transport schemes. }
  \begin{tabular}{ l | c | r }
    \hline
      & Central FV & UPW biased with FL \\\hline
    glevel 4 ($\sim 4^\circ$) & 300 s & 7200 s\\ \hline
    glevel 5 ($\sim 2^\circ$) & 150 s & 3600 s\\ \hline
    glevel 6 ($\sim 1^\circ$) & 75  s & 1800 s\\ \hline
    glevel 7 ($\sim 0.5^\circ$) & 37.5 s & 900 s\\\hline
  \end{tabular}
  \label{tab:testtimestep}
  \tablefoot{`Central FV' is the central finite-volume described in \cite{2001Tomita} and `UPW biased with FL' is the upwind-biased scheme from \cite{2007Miura} with the flux limiter from \cite{1996Thuburn}. The time step used for the upwind-biased simulations without the flux limiter is the same as using the flux limiter. We note that the number of steps for the full integration depends on the time step, which needs to be large enough to complete a full revolution across the sphere (e.g. 144 steps if the time step is 7200 s and if the period in Eq. \ref{eq:u0} is set to 12 days).}
\end{table}

\begin{figure*}
\begin{centering}
\subfigure[CFV]{
\includegraphics[width=0.65\columnwidth]{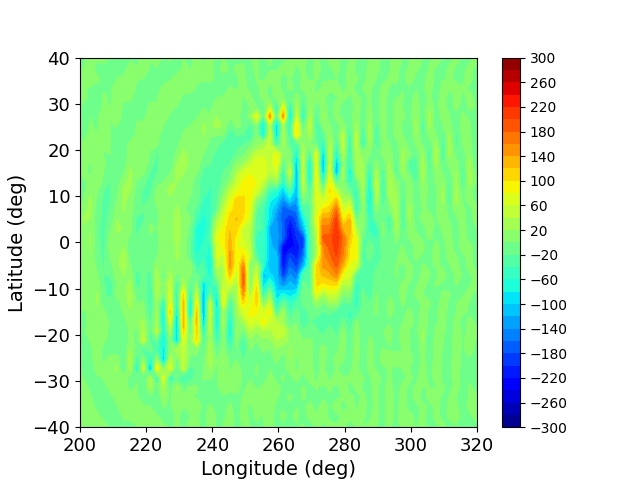}\label{fig:err_test_solversa}}
\subfigure[UW without FL]{
\includegraphics[width=0.65\columnwidth]{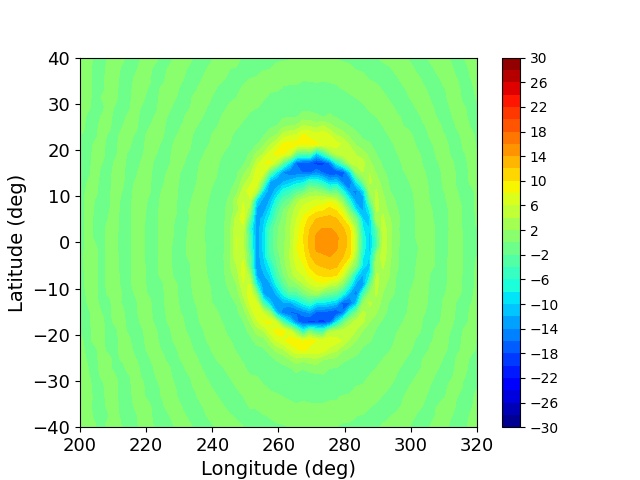}\label{fig:err_test_solversb}}
\subfigure[UW with FL]{
\includegraphics[width=0.65\columnwidth]{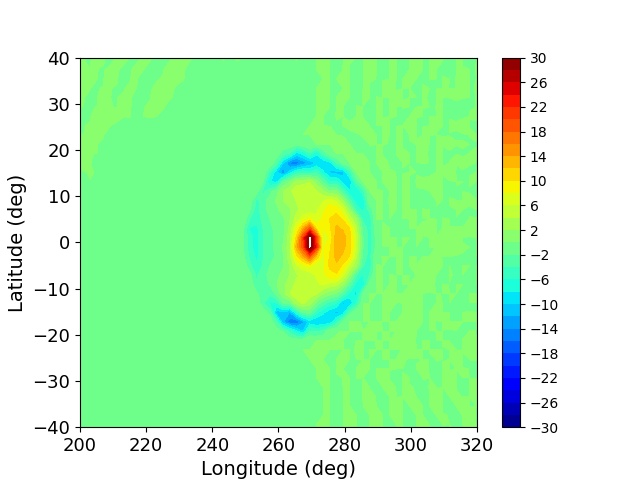}\label{fig:err_test_solversc}}
\caption{Latitude-longitude maps of the difference between the initial cosine bell and after a complete revolution across the sphere (spatial resolution glevel 6). The three maps refer to three different methods that used different time steps (Table \ref{tab:testtimestep}): CFV (central finite-volume), UW without FL (upwind biased without flux limiter), and UW with FL (upwind with flux limiter).}
\label{fig:err_test_solvers}
\end{centering}
\end{figure*}

For this simple test we implemented an explicit Euler time integration. The time-steps used for the different spatial resolutions and techniques are shown in Table \ref{tab:testtimestep}. The maps in Fig. \ref{fig:err_test_solvers} show the difference between the initial cosine bell and after a complete revolution using different methods. In Fig. \ref{fig:err_test_solversa}, the wavy pattern due to the dispersive nature of the central finite-volume method is very clear, despite using a much shorter time step than the one used by the other methods. The central finite-volume developed in \cite{2001Tomita} is not an upwind solver, which makes the solutions less diffusive and more dispersive. To mitigate the instabilities created by the central finite volume, we could add explicit diffusion to smear the wavy pattern. However, solving the diffusion terms reduces the solver's accuracy and makes it more time-consuming since it implies solving Laplacians in an icosahedral grid. The large amplitude of the error shows that this method is not stable without a diffusion term. Maps  \ref{fig:err_test_solversb} and  \ref{fig:err_test_solversc} show the results for the case of the upwind-biased scheme with and without the flux limiter, respectively. If the flux limiter is not used, the amplitude of the error is smaller than using the flux limiter. However, the error is not monotonically decreasing from the centre of the cosine bell distribution, which is an indication that the shape of the distribution is not well preserved. In \ref{fig:err_test_solversc}, the flux limiter is applied, and despite the error being relatively large at the centre of the cosine bell shape, the shape of the distribution is preserved. The larger error in \ref{fig:err_test_solversc} is related to the intrinsic diffusion of the flux-limiter scheme.

\begin{figure}[ht]
\includegraphics[width=1.0\columnwidth]{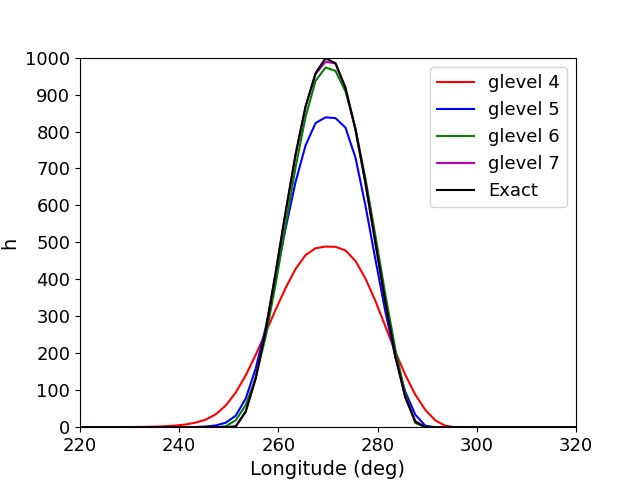}
\caption{Final tracer distributions at the equator from the cosine bell test case that uses the upwind-biased scheme with a flux limiter for different spatial resolutions. The solid black line is the analytical solution used for the initial condition (Eq. \ref{eq:cosinebell}).}
\label{fig:err_plot_diff}
\end{figure}

The plot in Fig. \ref{fig:err_plot_diff} shows the final equatorial tracer distribution for the simulations using the upwind-biased scheme with the flux limiter for different spatial resolutions.  We verify that by increasing the spatial resolution, the numerical diffusion becomes smaller and the numerical solutions are close to the exact solutions.

\begin{figure}[ht]
\includegraphics[width=1.0\columnwidth]{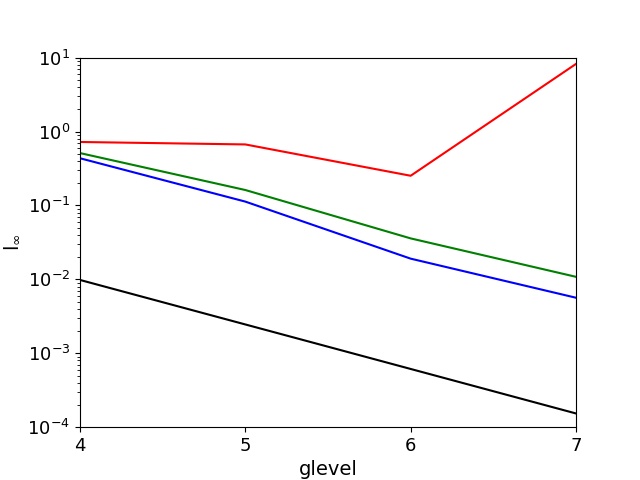}
\caption{$l_\infty$ convergence plot where glevel represents the different spatial resolutions. The various solid lines represent different methods: the central finite volume (red); the upwind-biased with flux limiter (green); and upwind biased without flux limiter (blue). The solid black line is a reference line with a slope of an exact second-order convergence.}
\label{fig:err_plot}
\end{figure}

To measure quantitatively the accuracy of our schemes, we computed the error, $l_\infty$, defined as
\begin{equation}
    l_{\infty} = max|h-h_{exact}|,
\end{equation}
where $h$ and $h_{exact}$ are the numerical integrated and the analytic solutions, respectively. $l_\infty$ is the largest error anywhere in our spherical grid. We consider $l_\infty$ to be the real error of the methods applied, which is not mitigated by an averaging method (e.g. root mean square error). In Fig. \ref{fig:err_plot} we show that the upwind-biased scheme improves the accuracy of the numerical solver considerably without the need for extra diffusion components. The slopes of the upwind schemes are similar to the exact second-order convergence. In some regions, the solver without using the flux limiter is even better than second-order accuracy.  For the central finite volume case we found as expected that the error increases with the spatial resolution, which is due to the dispersive nature of the solver. To improve the accuracy of using the central finite volume, as mentioned above, we can reduce the time step or apply explicit numerical diffusion.

\subsection{3D transport and consistency in the atmospheric mass transport}

The previous section described how the transport is calculated in the horizontal projection (parallel to the spherical surface). For the vertical integration, we used a second-order central finite difference approach with the flux limiter from \cite{1996Thuburn}.

We implemented the tracer transport in the 3D dynamical core. The time integration was done in the large time step integration section (see \cite{2016Mendoncab} for more information about \texttt{THOR}'s time integration). The tracer equation is integrated implicitly using the updated momenta values. We replaced the central finite volume with the upwind-biased scheme to solve the mass continuity equation to maintain consistency between the tracers and the total atmospheric mass. This way, we maintain consistency in the spatial distribution of the various types of mass transported in the model. Other quantities, such as the momenta and potential temperature, are still integrated with the central finite volume followed with explicit hyper-diffusion as explained in \cite{2016Mendoncab}. 

\section{Benchmark tests}
\label{sec:3dbenchmark_test}
\subsection{Simplified physics}

To test our new 3D formulation, we explored first the ocean Earth-like test case proposed in \cite{2016Thatcher}. The test is a modified version of the well-known Held-Suarez test (\citealt{1994Held}). The new test includes simplified moist processes, boundary-layer mixing, latent heat exchange, and sensible heat between the planet surface and atmosphere. The moist physics schemes included have been shown to represent quantitatively well the climate state of simulations with more complex physical methods (\citealt{2016Thatcher}).  In the section below, we summarise the main points of the physical schemes implemented from \cite{2016Thatcher} to make the reader more familiarised with the new routines implemented. More details on how to implement these schemes are available in \cite{2016Thatcher}. \cite{2016Thatcher} provides supplementary FORTRAN routines with the simplified physics implemented, which we converted to the CUDA C programming language and to work on our specific grid indexing. The inclusion of simplified physics allows us to focus on the performance of the new methods implemented in \texttt{OASIS} to represent mass transport in a moist atmosphere. 

It is well known that sudden adjustments of local atmospheric quantities due to moist processes can trigger undesirable large-scale gravity waves (\citealt{2016Thatcher}). The 3D numerical experiments proposed in this work allow us to study if our numerical schemes are robust enough to prevent undesirable large-scale gravity waves from being triggered under different planetary conditions (tidally and non-tidally locked configurations).

Using a similar approach, we adapted the routines used to study the ocean Earth-like planet to explore planets under a tidally locked configuration. The two test cases (ocean Earth-like planet and ocean tidally locked planets) proposed in this work allowed us to verify that the new \texttt{OASIS} can reproduce robust results under two very distinct planet scenarios where moist processes play an important role in the atmospheric dynamics.

Below, we describe briefly the physical schemes implemented in \texttt{OASIS} to reproduce the two test cases. The values for the physical parameters used in the simulations are shown in Table \ref{tab:planets}.

\subsubsection{Large-scale precipitation}
The physics modules for the benchmark test do not include a cloud scheme formulation. The gas that condenses in the atmosphere is removed instantaneously from the atmosphere without considering any atmospheric transport of the condensed material or re-evaporation processes. The temperature and specific humidity trends during condensation are defined as 
\begin{equation}
\frac{\partial T}{\partial t} = \frac{L}{c_p}C
\label{eq:dtdtC}
\end{equation}
\begin{equation}
\frac{\partial q}{\partial t} = -C
\label{eq:dqdt}
,\end{equation}
where $L$ is the latent heat of vaporisation at 0$^\circ$C, c$_p$ is the specific heat of the dry air, and $C$ is the condensation rate. The two physical trends defined in these two equations, Eqs. (\ref{eq:dqdt} and \ref{eq:dtdtC}), contribute to the physical trends in Eqs. \ref{eq:thor3}-\ref{eq:thor4} solved in the dynamical core. The condensation is defined by 
\begin{equation}
C=\frac{dq_{sat}}{dt}=\frac{1}{\Delta t}\Big(\frac{q-q_{sat}(T)}{1+\frac{L}{c_p}\frac{L\,q_{sat}}{R_q T^2}}\Big)
\label{eq:c}
,\end{equation}
where $\Delta t$ is the dynamical time step, $q$ is the water vapour mass mixing ratio, $T$ is the atmospheric temperature, $R_q$ is the gas constant of the condensible gas (water), and $q_{sat}$ is the saturation-specific humidity. The $q_{sat}$ was formulated following \cite{2012Reed},
\begin{equation}
q_{sat}(T)=\frac{\epsilon}{p}e_0^\star\exp\Big[-\frac{L}{R_q}\Big(\frac{1}{T}-\frac{1}{T_0}\Big)\Big]
\label{eq:qsat}
,\end{equation}
where $e_0^\star$ is the saturation vapour pressure at the control temperature $T_0$, $\epsilon$ is the ratio of the gas constant for dry air to that for water vapour.

If the relative humidity exceeds the saturation vapour pressure, latent heat is released and the condensate is immediately removed. This simple scheme allows us to calculate the large-scale precipitation rate ($P_{ls}$, \citealt{2016Thatcher}) from
\begin{equation}
P_{ls} \approx \frac{1}{\rho_{q}g}\sum_{k=1}^{nv}C_k(p_{k+1/2}-p_{k-1/2})
\label{eq:precipitation}
,\end{equation}
where $nv$ is the number of layers, $k$ the layer index, and $\rho_q$ is density of the condensible gas (water in our experiments) in the atmosphere.

\begin{table*}
  \begin{tabular}{ l | c | c | c }
    \hline
      & Earth-like Planet & Tidally Locked Planet & Units \\\hline
    \textbf{\texttt{OASIS} Model Parameters} & \multicolumn{3}{c}{ } \\ 
    Grid horizontal resolution & 1 & 1 & degrees\\ 
    Number of vertical layers & 39 & 39 & $\#$ \\ 
    Time step & 100 & 50 & s \\ 
    &  &  &  \\ 
    \textbf{Bulk Planet Parameters} & \multicolumn{3}{c}{ } \\ 
    Planetary radius, $R_{planet}$ & $6.371\times10^6$ & $6.371\times10^6$ & m\\
    Planetary rotation rate, $\Omega$ & $7.292\times10^{-5}$ & $1.212\times10^{-5}$ & s$^{-1}$ \\ 
    Surface gravity, g & 9.8 & 9.8 & m\,s$^{-2}$\\ 
    Top of the atmospheric domain, z$_{top}$ & 36 & 36 & km\\ 
    Specific heat of background gas at constant pressure, c$_p$ & 1004.6 & 1004.6 & J\,kg$^{-1}$\,K$^{-1}$\\
    Background gas constant, R$_d$ & 287.04 & 287.04 & J\,kg$^{-1}$\,K$^{-1}$\\
    Reference pressure, p$_{00}$ & 10$^5$ & 10$^5$ & Pa\\
    Planetary obliquity, $\epsilon$ & 0 & 0 & degrees\\
    Orbital eccentricity, $\alpha$ & 0 & 0 & -\\
    &  &  &  \\ 
    \textbf{Large-Scale Condensation} & \multicolumn{3}{c}{ } \\ 
     Ratio of gas constant for dry air to that for vapour, $\epsilon$ & 0.622 & 0.622  &  - \\ 
     Control temperature for q$_{sat}$ calculation, T$_0$ & 273.16 & 273.16 & K\\ 
     Saturation vapour pressure at T$_0$, $\epsilon_0^*$ & 610.78 & 610.78 & Pa \\ 
     Latent heat at T$_0$, L & 2.5$\times10^6$ & 2.5$\times10^6$ & J\,kg$^{-1}$\\ 
     Gas constant for water vapour, R$_q$ & 2.5$\times10^6$ & 2.5$\times10^6$ & J\,kg$^{-1}$K$^{-1}$\\
     Water density, $\rho_q$ & 1000 & 1000 & kg\,m$^{-3}$\\
     &  &  & \\ 
     \textbf{Surface conditions} & \multicolumn{3}{c}{ } \\
     Equator-to-pole difference, $\Delta$T & 29 & 100 & K\\
     SST at the pole, T$_{min}$ & 271 & 230 & K\\
     Width parameter for sea surface temperature, $\Delta l$ & 26 & 35 & degrees\\ 
      &  &  & \\ 
     \textbf{Surface Fluxes} & \multicolumn{3}{c}{ } \\
     Bulk transfer coefficients for sensible heat, C$_H$ & 0.0044 & 0.0044 & - \\
     Bulk transfer coefficients for water vapour, C$_E$ & 0.0044 & 0.0044 & - \\
     Constants for drag coefficient, C$_{d0}$ & - & 0.0007  & - \\
     Constants for drag coefficient, C$_{d1}$ & - & 0.000065 & - \\
     E-folding time for the Rayleigh friction, 1/k$_f$ & 1 &  -  & Earth day$^{-1}$\\
     Threshold sigma level for the Rayleigh friction, $\sigma_b$  & 0.7 & - & -\\ 
     Top of the boundary layer, p$_{pbl}$ & 850 & - & hPa\\
     Vertical mixing gradient parameter, p$_{strato}$ & 100 & 100 & hPa\\
     &  &  & \\ 
    \textbf{Radiation} & \multicolumn{3}{c}{ } \\ 
    Equator-pole temperature difference, $(\Delta T)_y$ & 65 & 40 & K\\
    Vertical temperature gradient parameter, $(\Delta\theta)_z$ & 10 & 40 & K\\
    Maximum equilibrium temperature, T$_{Equator}$ & 294 & 254 & K\\
    E-folding time parameter $\#1$ for temperature relaxation, 1/k$_a$ & 40 & 20 & Earth day$^{-1}$\\
    E-folding time parameter $\#2$ for temperature relaxation, 1/k$_s$ & 4 & - & Earth day$^{-1}$\\ 
    &  &  & \\ \hline
  \end{tabular}
  \caption{Main parameters for the 3D simulations.}
  \label{tab:planets}
\end{table*}

\subsubsection{Boundary conditions}

In both simulations explored, we prescribed the surface temperature of the ocean covered planets. The formulation for the surface temperature in the ocean Earth-like simulation is given in \cite{2012Reed},
\begin{equation}
T_s=\Delta T \exp\Big(-\frac{\theta^2}{2(\Delta\,l)^2}\Big)+T_{min}
\label{eq:ts}
,\end{equation}
where $T_{min}$ is minimum temperature at the poles, $\Delta\,T$ is the temperature difference between equator and poles, $\theta$ is the latitude and $\Delta\,l$ is the width of the Gaussian function in Eq. \ref{eq:ts}.

For the tidally locked planet experiment, we modified Eq. \ref{eq:ts} to represent the day-night contrast,
\begin{equation}
T_s=\Delta T \exp\Big(-\frac{\theta^2}{2(\Delta\,l)^2}\Big)\exp\Big(-\frac{\phi^2}{2(\Delta\,l)^2}\Big)+T_{min}
\label{eq:ts_2}
,\end{equation}
where $\phi$ is the longitude. The parameters for the tidally locked planet were calibrated to the results of \cite{2019Yang}.

\subsubsection{Surface fluxes}
To represent the mechanical interaction between the atmosphere and the surface we implemented two different schemes for the ocean Earth-like simulation and the tidally locked planet. For the ocean Earth-like simulation, we implemented the Rayleigh linear friction scheme exactly as done in the Held-Suarez experiment (\citealt{1994Held}),
\begin{equation}
    \frac{\partial\textbf{v}_h}{\partial t} = K_v(\sigma)\textbf{v}_h,
    \label{eq:srf_flx_dvdt}
\end{equation}
\begin{equation}
    K_v(\sigma) = k_f\times\max\Big(0.0,\frac{\sigma - \sigma_b}{1-\sigma_b}\Big),
    \label{eq:srf_flx_kv}
\end{equation}
where $\textbf{v}_h$ is the 3D wind field projected parallel to the spherical surface, $K_v$ is the strength of the frictional damping, $\sigma_b$ sets the depth of the winds damped, and $k_f$ is the strength of the surface dissipation. The values for the parameters used in this scheme are calibrated to do Earth-like simulations. \cite{2016Thatcher} used the same scheme and constant values. In the tidally locked planet, we did one further step in terms of complexity to allow the code to be more flexible but still trying to keep it simple in terms of implementation. We followed the implementation suggested in \cite{2012Reed}. The time rate of each horizontal velocity component in the lowermost layer is defined as
\begin{equation}
    \frac{\partial u_a}{\partial t} = -\frac{C_d(\sqrt{u_a^2+v_a^2})u_a}{z_a},
    \label{eq:u_a}
\end{equation}
\begin{equation}
    \frac{\partial v_a}{\partial t} = -\frac{C_d(\sqrt{u_a^2+v_a^2})v_a}{z_a},
    \label{eq:v_a}
\end{equation}
where $u_a$ and $v_a$ are the horizontal components of the wind at the lowermost model layer, $C_d$ is the drag coefficient and $z_a$ is the altitude of the first layer ($z_a$ is set to 50 m in both simulations). We are assuming that the wind velocities at the ocean surface are zero. The drag coefficient, $C_d$, depends on the magnitude of the horizontal wind at the lowermost layer as
\begin{equation}
    C_d = \begin{cases}C_{d0}+C_{d1}(\sqrt{u_a^2+v_a^2}) &\text{if $\sqrt{u_a^2+v_a^2} < 20 $m\,s$^{-1}$}\\
0.002 &\text{if $\sqrt{u_a^2+v_a^2} \ge 20 $m\,s$^{-1}$}.
    \end{cases}
\label{eq:cd}
\end{equation}
The constants $C_{d0}$ and $C_{d1}$ are the same as in \cite{2008Smith}. The vertical velocity time rate is set to zero. This simple formulation represents the eddy turbulence caused by the interaction between ocean surface and atmosphere. The formulation for the kinematic eddy flux of water vapour and heat at the surface is the same for both simulations:
\begin{equation}
    \frac{\partial q_a}{\partial t} = -\frac{C_H(\sqrt{u_a^2+v_a^2})(T_s-T_a)}{z_a},
    \label{eq:T_a}
\end{equation}
\begin{equation}
    \frac{\partial T_a}{\partial t} = -\frac{C_E(\sqrt{u_a^2+v_a^2})(q_{sat,s}-q_a)}{z_a},
    \label{eq:q_a}
\end{equation}
where $C_E$ and $C_H$ are the bulk transfer coefficients for water and heat, respectively. $T_s$ is the surface temperature, and $q_{sat,s}$ is the saturation specific humidity calculated at surface temperature. The bulk transfer coefficients were set to 0.0044, which are four times higher than as in \cite{2008Smith}, as suggested in \cite{2016Thatcher}. More details about the scheme are available in \cite{2012Reed}.

\subsubsection{Boundary layer}
In the ocean Earth-like simulation, we applied the Rayleigh friction from \cite{1994Held} to represent the boundary-layer mixing of the horizontal winds. This simplification was applied to be consistent with the results from \cite{2016Thatcher}. For the ocean tidally locked planet simulation, we used a simple diffusive boundary scheme. This scheme is an extension of the formulation described in the section above for the surface fluxes. The turbulent kinematic mixing is represented by 
\begin{equation}
    \frac{\partial u}{\partial t} = - \frac{1}{\rho}\frac{\partial \rho \overline{w'u'}}{\partial z} = \frac{1}{\rho}\frac{\partial} {\partial z} \rho K_m \frac{\partial u} {\partial z}
    \label{eq:blu}
\end{equation}
\begin{equation}
    \frac{\partial v}{\partial t} = - \frac{1}{\rho}\frac{\partial \rho \overline{w'v'}}{\partial z} = \frac{1}{\rho}\frac{\partial} {\partial z} \rho K_m \frac{\partial v} {\partial z}
    \label{eq:blv}
,\end{equation}
where the prime symbols represent the deviations from time-averaged quantities, the overbar is the time average operator, $w$ is the vertical velocity, and $K_m$ is the eddy diffusivity coefficient for momentum. The eddy coefficient $K_m$ is defined as
\begin{equation}
    K_m = \begin{cases}C_d(\sqrt{u_a^2+v_a^2})z_a &\text{if $p$ < $p_{top}$}\\
C_d(\sqrt{u_a^2+v_a^2})z_a\exp\Big(-\Big[\frac{p_{top}-p}{p_{strato}}\Big]^2\Big) &\text{if $p \le p_{top}$}.
    \end{cases}
\label{eq:Km}
\end{equation}
The variable $p_{top}$ sets the top of the boundary layer, which for simplicity is set at the same number for every column. For altitudes above $p_{top}$, the strength of $K_m$ decreases exponentially with a rate set by $p_{strato}$.

The vertical turbulent mixing of potential temperature and specific humidity is the same in both simulations, 
\begin{equation}
    \frac{\partial \Theta}{\partial t} = - \frac{1}{\rho}\frac{\partial \rho \overline{w'\Theta'}}{\partial z} = \frac{1}{\rho}\frac{\partial} {\partial z} \rho K_E \frac{\partial \Theta} {\partial z}
    \label{eq:blth}
\end{equation}
\begin{equation}
    \frac{\partial q}{\partial t} = - \frac{1}{\rho}\frac{\partial \rho \overline{w'q'}}{\partial z} = \frac{1}{\rho}\frac{\partial} {\partial z} \rho K_E \frac{\partial q} {\partial z}
    \label{eq:blq}
,\end{equation}
where $K_E$ is the eddy diffusivity coefficient for energy defined as
\begin{equation}
    K_E = \begin{cases}C_E(\sqrt{u_a^2+v_a^2})z_a &\text{if $p$ < $p_{top}$}\\
C_E(\sqrt{u_a^2+v_a^2})z_a\exp\Big(-\Big[\frac{p_{top}-p}{p_{strato}}\Big]^2\Big) &\text{if $p \le p_{top}$}.
    \end{cases}
\label{eq:KE}
\end{equation}
The formulation of $K_E$ is identical to $K_m$. More details about this simple boundary layer scheme can be found in \cite{2012Reed} and \cite{2016Thatcher}.

\subsubsection{Radiation}

A Newtonian temperature relaxation represents the impact of the radiation in the atmosphere. The scheme is very similar for the two planets explored. The diabatic heating for the ocean Earth-like experiment is almost identical to the Held-Suarez test (\citealt{1994Held}). The temperatures in the atmosphere are forced towards a radiative-convective equilibrium temperature ($T_{eq}$) at a specific timescale ($k_T$): 
\begin{equation}
\frac{\partial T}{\partial t} = -k_T(\phi,\sigma)\Big[T-T_{eq}(\phi,\sigma)\Big].
\label{eq:dtdt}
\end{equation}
The equilibrium temperature is defined such as in \cite{1994Held},
\begin{align}
\begin{split}
T_{eq}(\phi, p)= &max\Big\{200 K, \Big[T_{Equator} - (\Delta T)_y\sin^2\phi -\\
&(\Delta\theta)_z log\Big(\frac{p}{p_{00}}\Big)\cos^2\phi\Big]\Big(\frac{p}{p_{00}}\Big)^\kappa\Big\}.
\end{split}
\label{eq:teq}
\end{align}
The values of each parameter are defined in Table \ref{tab:planets}. The differences between values used in our simulations (suggested in \citealt{2016Thatcher}) and the original Held-Suarez test (\citealt{1994Held}) are the values for $T_{Equator}$ and $(\Delta T)_y$. In \cite{2016Thatcher}, $T_{Equator}$ and $(\Delta T)_y$ are 315\,K and 65\,K, respectively, instead of 294\,K and 60\,K from \cite{1994Held}.

The timescale for the temperature forcing is defined as
\begin{equation}
k_T(\phi,\sigma) = k_a + (k_s - k_a)max\Big(0, \frac{\frac{p}{p_{00}}-\sigma_b}{1-\sigma_b}\Big)\cos^4\phi.
\label{eq:kt}
\end{equation}
\newline
The different parameters in the equation are constants defined in Table \ref{tab:planets}. To represent the radiative-convective forcing in the tidally locked planet, we changed both Eqs. \ref{eq:teq} and \ref{eq:kt}. The equations and the parameters were adapted to reproduce a similar temperature structure to the case of a tidally locked planet orbiting an M-star from \citealt{2019Yang}. Other works have also used Newtonian temperature relaxation to explore the climate of terrestrial tidally locked planets, such as \cite{2011Heng} and \cite{2015Carone}. Our modified equations for this experiment are
\begin{align}
\begin{split}
T_{eq}(\phi, p)= &max\Big\{200 K, \Big[T_{Equator} - \cos\theta_z((\Delta T)_y -\\
&(\Delta\theta)_z log\Big(\frac{p}{p_{00}}\Big))\Big]\Big(\frac{p}{p_{00}}\Big)^\kappa\Big\},
\end{split}
\label{eq:teq_2}
\end{align}
\begin{equation}
k_T(\phi,\sigma) = k_a.
\label{eq:kt_2}
\end{equation}
The relaxation time coefficient is set as a constant in the tidally locked planet simulation. In Eq. \ref{eq:teq_2}, the term $\cos_z$ refers to the cosine of the zenith angle. We again note that the value of each parameter is explicitly shown in Table \ref{tab:planets}.

\begin{figure}[ht]
\includegraphics[width=1.0\columnwidth]{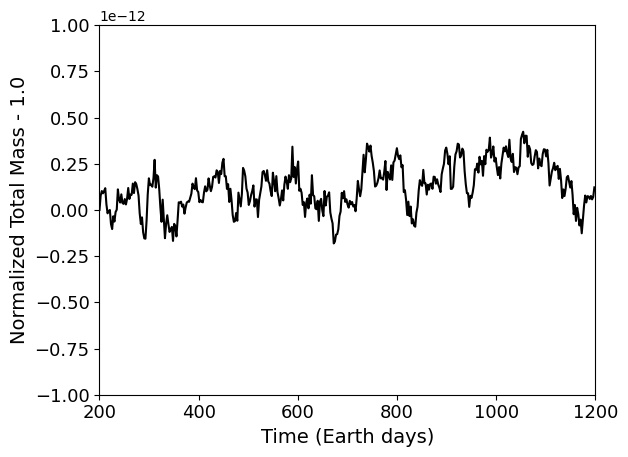}
\caption{Evolution of the total mass error in the ocean Earth-like simulation.}
\label{fig:mass_conservation}
\end{figure}

\begin{figure*}
\begin{centering}
\subfigure[Zonal winds]{
\includegraphics[width=0.9\columnwidth]{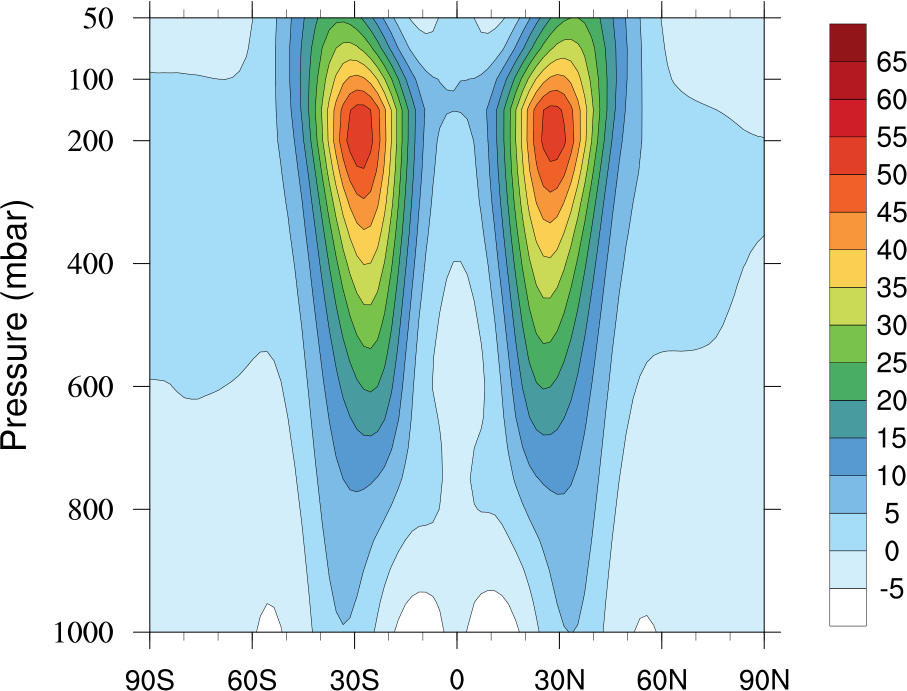}\label{fig:zwinds_temp_eartha}}
\hspace{0.5cm}
\subfigure[Temperature]{
\includegraphics[width=0.9\columnwidth]{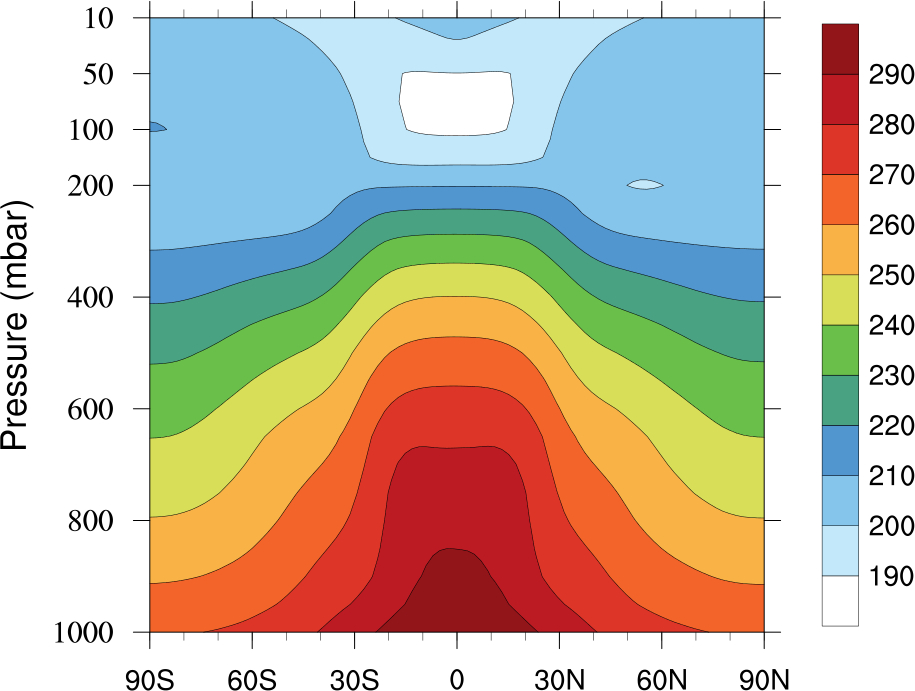}\label{fig:zwinds_temp_earthb}}
\caption{Final time- and longitudinal-averaged zonal winds (m$\,$s$^{-1}$) and temperatures (K) for the ocean Earth-like experiment. The values were time-averaged for 1000 Earth days.}
\label{fig:zwinds_temp_earth}
\end{centering}
\end{figure*}

\section{Results from 3D simulations}
\label{sec:results_3d}
This section presents the 3D simulations from the Earth-like planet and the tidally locked planet, where both have the surface covered with an ocean. As explained in the previous section, these simulations include simplified physics schemes and aim to reproduce conditions that allow us to evaluate the mass and moist transport accuracy in 3D simulations. Models containing poor representations of moist transport in the atmosphere will produce errors that propagate, for example, to the energy budget of the atmosphere, cloud cover, and atmospheric circulation, which can compromise the interpretation of observational data on exoplanets. To assess the simulations, we characterise the global circulation in both simulations and pay particular attention to possible sources of high-frequency waves associated with bad numerical implementations. For the ocean Earth-like simulation, \cite{2016Thatcher} showed that the simplified physics applied here can reproduce the results from models that used more complex physics formulations (e.g. including microphysics). In the section below, we compare our results to the results presented in \cite{2016Thatcher}. We suggest that the results obtained in this work become the first step towards a complete benchmarking of future moist atmospheric simulators on terrestrial planets.  

\subsection{Ocean Earth-like planet}
\label{subsec:AquaEarth}
We integrated our model for 1200 Earth days with a time step of 100 seconds. The first 200 Earth days are discarded as being part of the numerical spin-up phase of the simulation. 

The new solvers implemented in \texttt{OASIS} aim to improve the mass transport representation in the 3D simulations. Two of the main improvements are the shape-preserving of the transported mass distribution and mass conservation. Figure \ref{fig:mass_conservation} shows how the total mass error ($M_{error}$) evolved during the simulation. The total mass error was calculated using the following equation,
\begin{equation}
    M_{error} = \frac{M}{M_i} - 1,
    \label{eq:error_M}
\end{equation}
where $M$ is the spatially integrated atmospheric mass for a particular instant and $M_i$ is the spatially integrated atmospheric mass at the beginning of the simulations. The small error measured along the simulation is always below 5$\times$10$^{-13}\%$, and it does not present any constant drift, which highlights the robustness of the new formulation.

\subsubsection{Zonal winds and temperature}
The spin-up of the simulated atmosphere is mainly driven by the thermal forcing and the rotational effects. In Fig. \ref{fig:zwinds_temp_earth} we show the longitudinal- and time-averaged temperature and zonal\footnote{Longitudinal component of the wind velocity.} winds.  As expected, the temperatures are warmer at low latitudes, where the lapse rate\footnote{Vertical temperature gradient.} is also larger. The tropopause is located at roughly 200 mbar. The cold and inactive stratosphere is located above the 200 mbar. Hotter layers in the atmosphere extend to higher altitudes compared to the traditional Held-Suarez Test (\citealt{1994Held}), which is driven by the surplus of heat provided from the condensation in the large-scale precipitation scheme. The more significant latitudinal gradients in temperature drive the formation of stronger winds, as it is possible to derive from the approximated geostrophic thermal wind equation. The average longitudinal winds in our simulation reach a maximum of around 50-55 m$\,$s$^{-1}$ at 200 mbar and 30$^\circ$ latitude. The winds decrease their magnitude in general with the decrease in altitude. Our winds and temperature results are quantitatively similar to the results presented in \cite{2016Thatcher}. However, the mid-latitude jets in our simulation are slightly weaker (roughly 5 m$\,$s$^{-1}$ difference). The differences are likely caused by the different levels of diffusion applied in the various models or different model implementations, such as our model diagnostics being updated in the dynamical core instead of being done immediately in the physical core (\citealt{2020Mendonca}; \citealt{2020Deitrick}). 

\begin{figure}[ht]
\includegraphics[width=1.0\columnwidth]{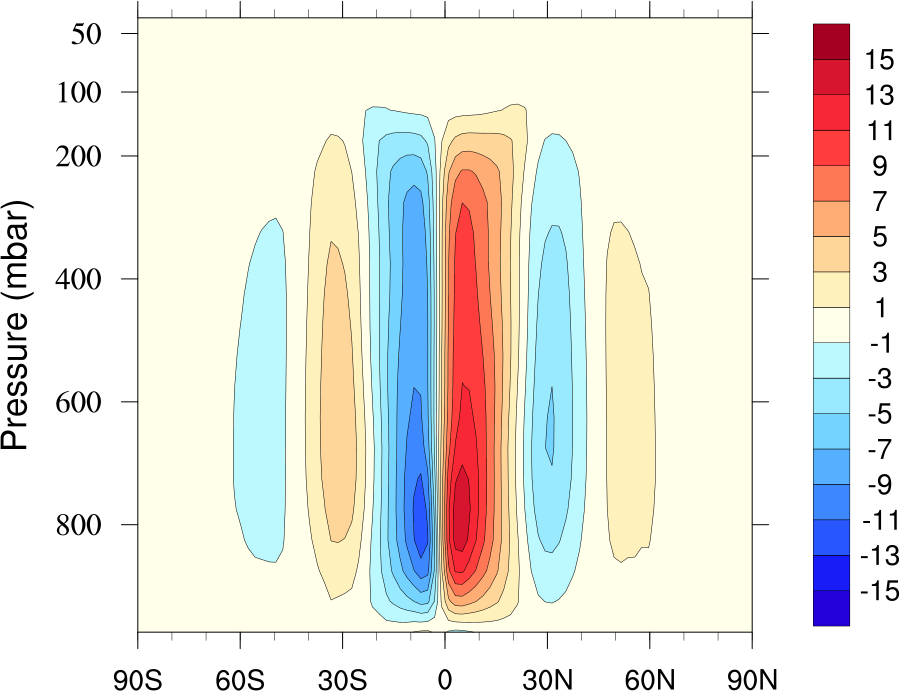}
\caption{Final time- and longitudinal-averaged mass stream function function (10$^{10}$kg$\,$s$^{-1}$) for the ocean Earth-like experiment. The values were time-averaged for 1000 Earth days.}
\label{fig:psi_earth}
\end{figure}

\begin{figure*}
\begin{centering}
\subfigure[Meridional heat flux]{
\includegraphics[width=0.9\columnwidth]{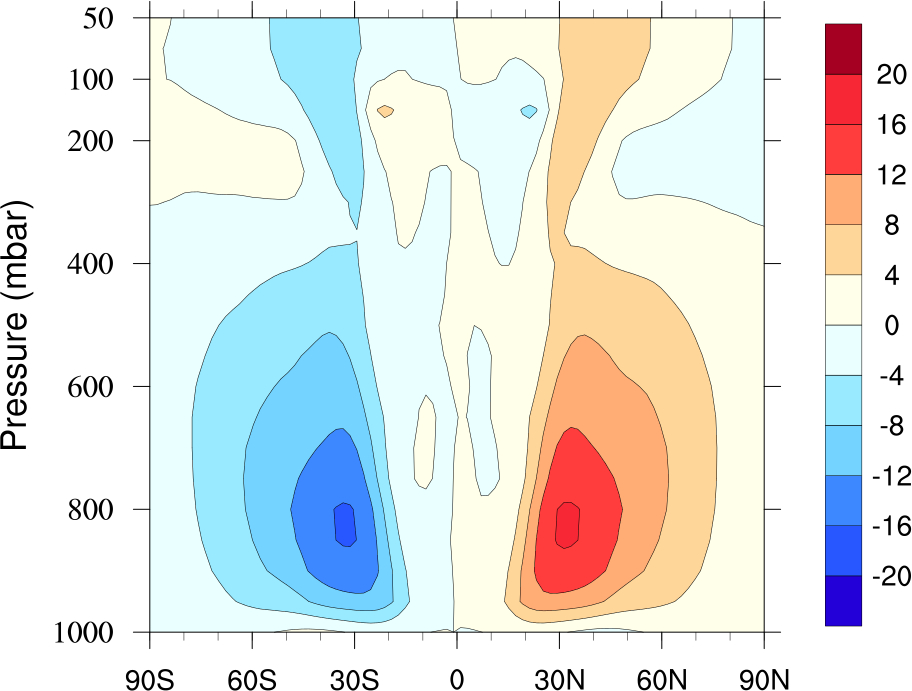}\label{fig:mhf_eke_eartha}}
\hspace{0.5cm}
\subfigure[Eddy kinetic energy]{
\includegraphics[width=0.9\columnwidth]{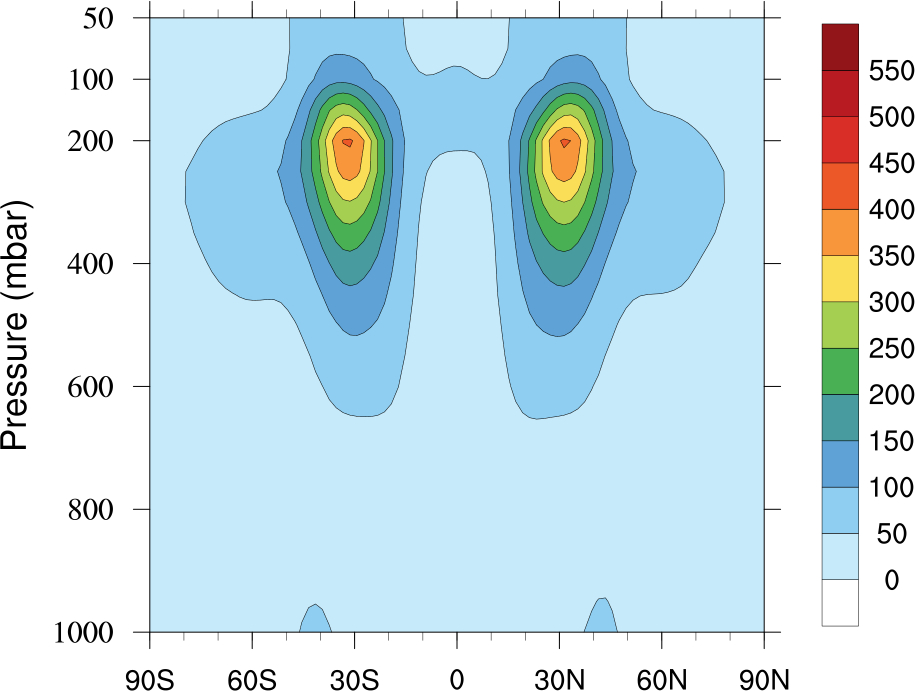}\label{fig:mhf_eke_earthb}}
\caption{Final time- and longitudinal-averaged meridional eddy sensible heat flux (m$\,$s$^{-1}\,$K) and eddy kinetic energy (m$^2\,$s$^{-2}\,$K) for the ocean Earth-like experiment. The values were time-averaged for 1000 Earth days.}
\label{fig:mhf_eke_earth}
\end{centering}
\end{figure*}

\subsubsection{Mass stream function}
In Fig. \ref{fig:psi_earth} we show the mass stream function, which we calculated from the following equation:
\begin{equation}
    \Psi = \frac{2\pi R_{planet} \cos\phi}{g}\int_0^P\overline{[v]}dp,
\end{equation}
where $R_{planet}$ is the planet radius, $\phi$ is the latitude, $g$ is the surface gravity, $p$ is the atmospheric pressure and $\overline{[v]}$ is the zonal and time-averaged meridional wind\footnote{Latitudinal component of the wind velocity.}. The results show the well-known atmospheric cells: Hadley, Ferrel, and polar cells (latitudes higher than 50$^\circ$. The latitudinal extension of the atmospheric cells is smaller in this test than in the Held-Suarez test, which assumes a dry atmosphere. The characteristics of the atmospheric cells simulated in this work are consistent with the results from, e.g. \cite{2008LeeM-I} and \cite{2016Thatcher}. The different positions of the atmospheric cells also impact the latitudinal position of the mid-latitudinal jets that are moved into lower latitudes. The stronger up-welling motion by mean circulation is located in the equatorial region at pressure levels around 800 mbar, where large-scale precipitation is at its maximum. 

\subsubsection{Meridional eddy sensible heat flux and eddy kinetic energy}
Wave activity has an important role in driving atmospheric circulation in the ocean Earth-like planet. In Fig. \ref{fig:mhf_eke_earth} we show the meridional eddy sensible heat flux and the eddy kinetic energy for the ocean Earth-like simulation. These two quantities are calculated from the eddy components\footnote{Zonal or longitudinal inhomogeneities.} of the temperature and wind field. In general, atmospheric waves transport sensible heat from the tropics to higher latitudes. The meridional eddy sensible heat flux ($\overline{[v'T']}$) map shows a dominant poleward eddy heat transport below the pressure level 400 mbar. The poleward transport of eddy sensible heat flux becomes again more intense above 200 mbar. The maximum of meridional eddy sensible heat flux is located between 30$^\circ$ to 40$^\circ$ latitude at roughly 850 mbar. The eddy kinetic energy (0.5$\overline{[u'u' + v'v'+ w'w']}$) shown in Fig. \ref{fig:mhf_eke_earth}, reaches the maximum at the mid-latitude jet's location, where the largest dynamical wave-activity is present. Both meridional heat flux and eddy kinetic energy obtained here are very similar to the results from the dry atmosphere Held-Suarez experiment (\citealt{1994Held} and \citealt{2016Mendoncab}).

\subsubsection{Vertical velocity}
\begin{figure}[ht]
\includegraphics[width=1.0\columnwidth]{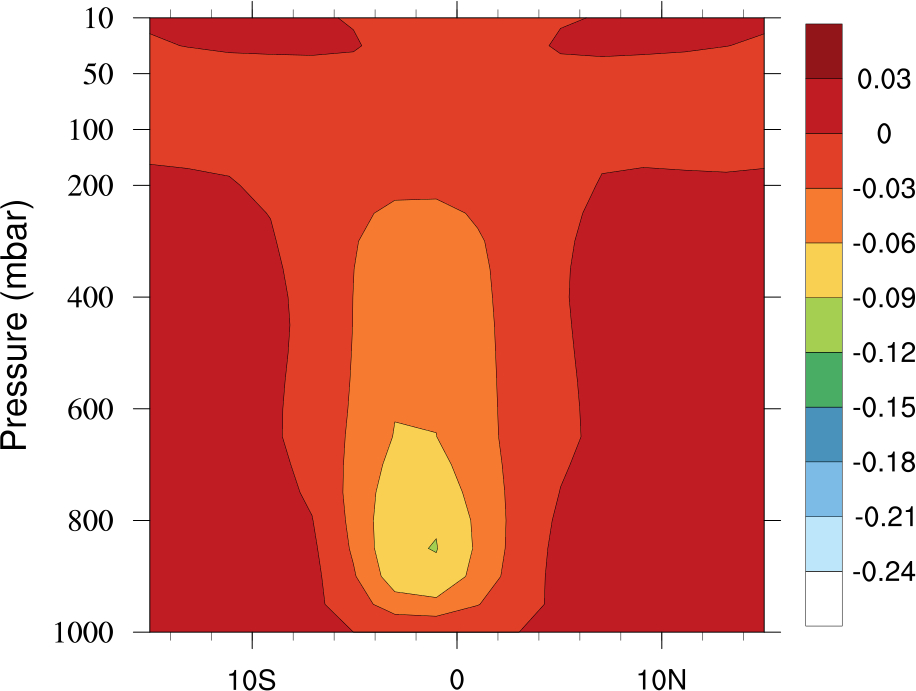}
\caption{Final time- and longitudinal-averaged vertical velocity (Pa$\,$s$^{-1}$) between 15$^\circ\,$S and 15$^\circ\,$N for the ocean Earth-like simulation. The values were time-averaged for 1000 Earth days.}
\label{fig:w_earth}
\end{figure}

Figure \ref{fig:w_earth} shows the vertical velocity in the tropical region. The vertical velocity results are consistent with the mass-stream function shown above. The negative values near the equator represent the upward branch of the Hadley cell, which reverses its direction for latitudes above roughly 6$^\circ$. The strength of the Hadley circulation is enforced by the condensation of the water vapour (release of latent heat), which is more abundant in the equatorial region.

\begin{figure}[ht]
\includegraphics[width=1.0\columnwidth]{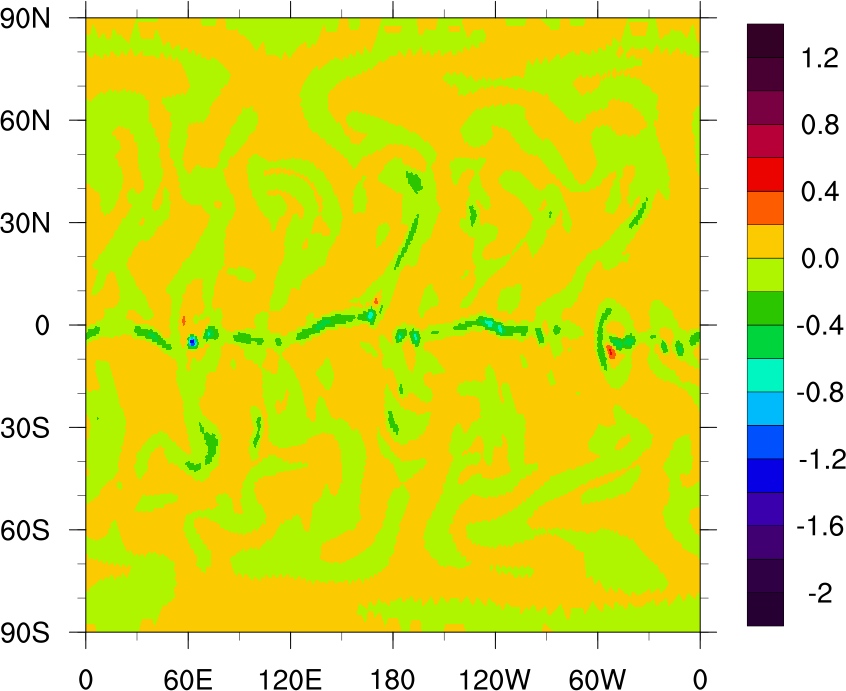}
\caption{Instantaneous latitude-longitude map of vertical velocity (Pa$\,$s$^{-1}$) for the ocean Earth-like experiment.}
\label{fig:w_inst_earth}
\end{figure}

In Fig. \ref{fig:w_inst_earth} we show a snapshot of the vertical velocity at a pressure level close to the surface. The predominant negative value in the equatorial region is related to the previously discussed upward branch of the Hadley cell. It is also clear from the patterns in the vertical velocity map that the wave activity propagates from the tropical region towards higher latitudes. The vertical winds near the surface are also important to check for any numerical noise associated with the underlying grid (grid imprinting). A small perturbation in the vertical wind field would be easily detected in a simulation due to its low magnitude compared to other diagnostic variables (\citealt{2012Ullrich}). In our model, we use the icosahedral grid. If the numerical noise becomes significant, we would see the geometrical structure of the grid, which is not the case in our simulations. In regions of a large gradient of the model diagnostics, our dispersive finite-volume methods, such as the one we apply to the momentum and entropy, can produce numerical noise, which would be easily seen as high gravity waves. However, our solver is robust and did not show any sign of numerical problems associated with the nature of the central finite volume.

\subsubsection{Kinetic energy spectrum}
\begin{figure}[ht]
\includegraphics[width=1.0\columnwidth]{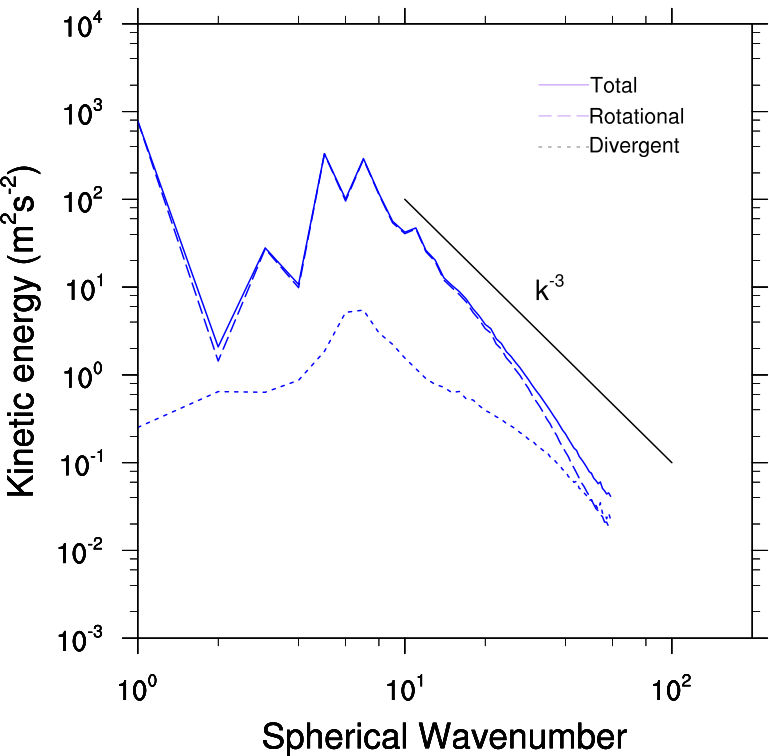}
\caption{Time-averaged kinetic energy spectrum at 0.25 mbar for the ocean Earth-like simulation. The dashed and dotted lines are the rotational and divergent components of the kinetic energy, respectively, and the solid line is the total (sum of the rotational and divergent components).}
\label{fig:kspectrum_earth}
\end{figure}

Kinetic energy spectra provide important information about how atmospheric motion is redistributed across different spatial scales. In Fig. \ref{fig:kspectrum_earth} we show the rotational and divergence components of the kinetic energy as a function of spherical wavenumber (e.g. \cite{2013Augier}). Our results show that at large scales (low wavenumbers), the rotational component is on average orders of magnitude larger than the divergent component. The magnitude of the divergent component increases on average for smaller horizontal scales, where for example, diabatic heating can increase baroclinicity and intensify thermally direct circulation. The two components become comparable at wavenumbers close to the truncation scales, where the energy in the divergent component becomes similar to the energy of the rotational component. In Fig. \ref{fig:kspectrum_earth}, we include a reference line of $k^{-3}$, where $k$ represents the spherical wavenumbers. The slope results from the downscale cascade of the enstrophy, which has been suggested by theoretical work and observations (e.g. \citealt{1985Nastrom}). At higher resolution simulations, it would be possible to capture the transition to $k^{-\frac{5}{3}}$ as a result of the downscale cascade of energy (\citealt{2000Lindborg}). Despite the simple physics represented in our simulation, our results are broadly consistent with previous studies based on observational or reanalysis datasets of Earth's atmosphere (e.g. \citealt{1999Koshyk}). The model needs to remove enstrophy or energy down at the spectral truncation scale to reproduce physically consistent results. Otherwise, physical quantities would build up at the truncation resolution. It is common practice for large-scale atmospheric models to remove kinetic energy near the truncation limit via, for example, explicit hyper-diffusion methods or numerical implicit schemes. Our spectrum in Fig. \ref{fig:kspectrum_earth} shows a steeper slope than $k^{-3}$ at higher wavenumbers as expected by the use of the fourth-order hyper-diffusion and divergence damping (\citealt{2016Mendoncab}). In general, simulations should overestimate the dissipation near the truncation resolution to maintain good stability performance and avoid non-physical artefacts arising from the sudden cut in resolution aliasing onto the well-resolved modes (smaller wavenumbers). Also, the physics at the truncation scales are not accurately solved by the atmospheric models. For an accurate representation of the large-scale physical phenomena, we should mitigate the inaccuracies by removing the energy at these scales.

\subsubsection{Water vapour concentration, relative humidity, and precipitation}

\begin{figure*}
\begin{centering}
\subfigure[Water vapour concentration]{
\includegraphics[width=0.9\columnwidth]{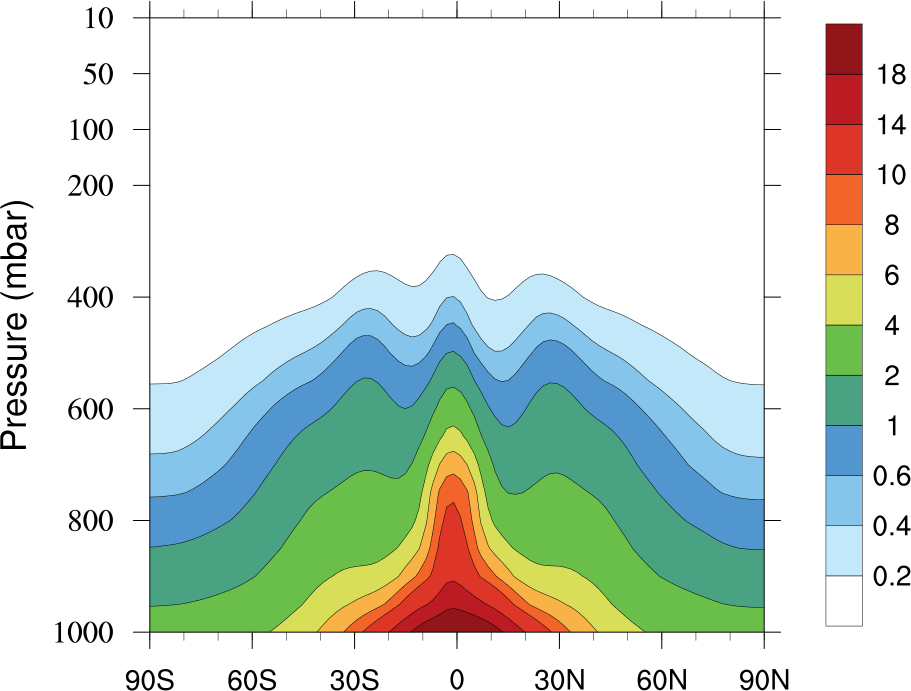}\label{fig:qwater_earth}}
\hspace{0.5cm}
\subfigure[Relative humidity]{
\includegraphics[width=0.9\columnwidth]{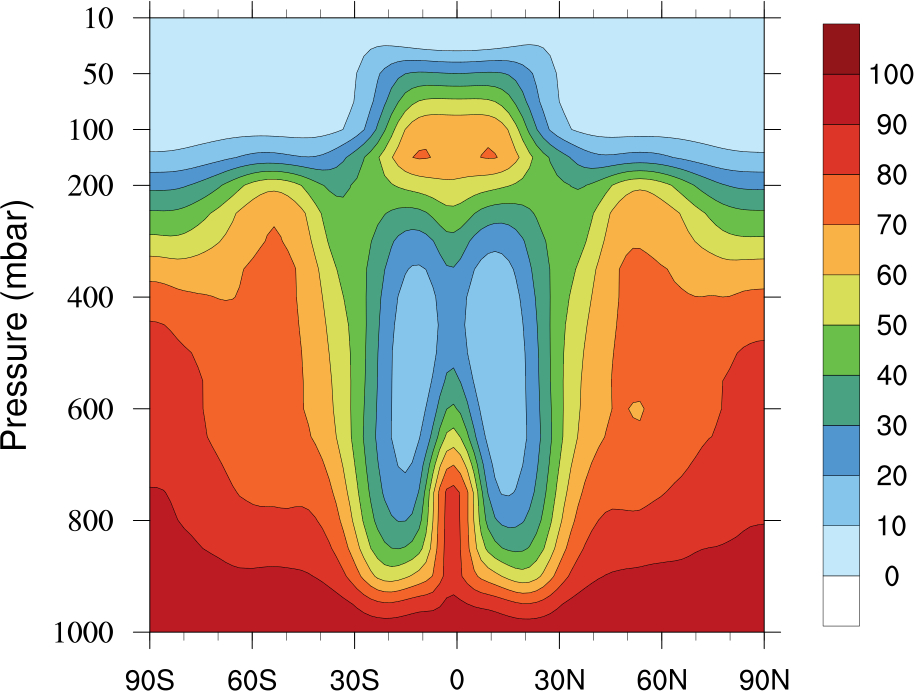}\label{fig:relvh_earth}}
\caption{Final time- and longitudinal-averaged water vapour concentration (kg/kg) and relative humidity ($\%$) for the ocean Earth-like simulation. The values were time-averaged for 1000 Earth days.}
\label{fig:q_relh_earth}
\end{centering}
\end{figure*}

In Fig. \ref{fig:qwater_earth} we show how the time and longitudinal average water vapour concentration is distributed across latitudes and pressures. The water vapour is evaporated from the ocean surface, predominantly from the equatorial region. The larger concentration of water vapour remains at low latitude, which is also the warmer atmospheric region, as we showed in Fig. \ref{fig:zwinds_temp_earth}. The elongated vertical feature in the equatorial region is driven by the accelerated vertical winds caused by the latent heat released by water condensation. Water vapour saturation is higher in the lower atmosphere, as is shown in Fig. \ref{fig:relvh_earth}. The low saturation levels at low latitudes are caused by large-scale precipitation.

\begin{figure}[ht]
\includegraphics[width=1.0\columnwidth]{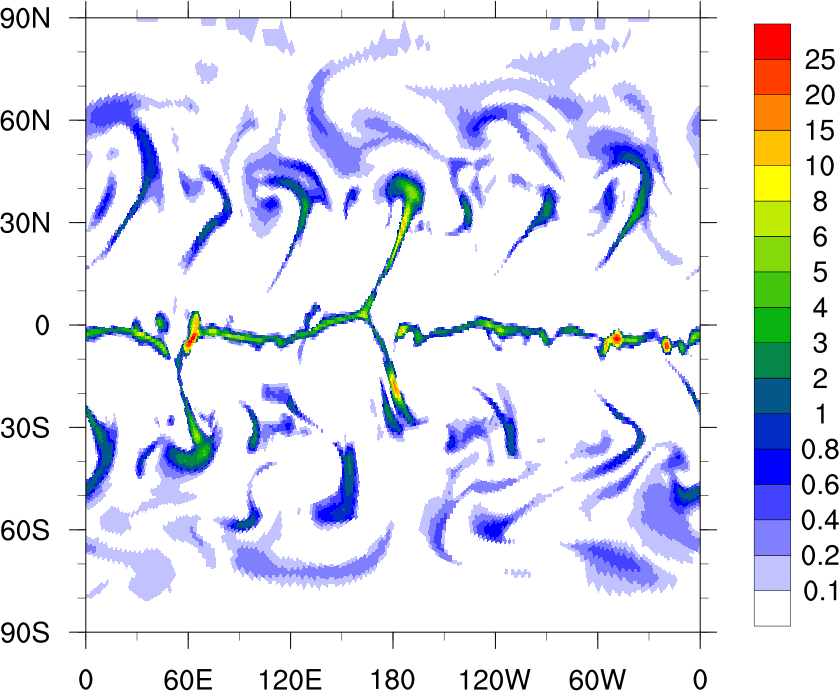}
\caption{Instantaneous latitude-longitude map of surface precipitation (mm$\,$h$^{-1}$) for the ocean Earth-like simulation.}
\label{fig:percipitation_earth}
\end{figure}

In Fig. \ref{fig:percipitation_earth} we plot the simulated surface precipitation. As expected from the distribution of the water vapour concentration, the higher levels of precipitation are located in the equatorial region. Despite the complex distribution of rainfall across the surface, the pattern obtained (climate state) is persistent. At mid-latitudes, instabilities grow by feeding on the available potential energy associated with the meridional temperature gradient (baroclinic instabilities). These instabilities transport higher levels of precipitation to higher latitudes. However, the region between 15–20$^\circ$ latitude, shows very low levels of precipitation. This complex system of precipitation is also obtained in simulations by complex Earth climate models (e.g. including cloud microphysics) on ocean Earth-like simulations (e.g. \citealt{2016Thatcher}).

The absence of any sign of numerical error that could compromise the physics of the simulations and the quantitatively very similar results between our simulation and \cite{2016Thatcher} demonstrates that our new \texttt{OASIS} has passed a critical ocean Earth-like benchmark test. 

\subsection{Ocean tidally locked planet}
\label{subsec:AquaExoplanet}
In this section we analyse the results from an ocean covered tidally locked planet experiment. The model parameters are estimated to reproduce quantitatively the atmospheric (plus ocean) temperature structure of the tidally locked planet orbiting an M star presented in \cite{2019Yang}. The planet parameters for our simulations are shown in Table \ref{tab:planets}. The simulations in \cite{2019Yang} are used as a reference, and we did not try to reproduce the same results since we use simpler physical numerical schemes in our simulations (e.g. we do not include radiative feedbacks from clouds). Our primary focus is to build an easy setup simulation to be used in the future as a benchmark test for moist mass transport in tidally locked planets. Note that in \cite{2019Yang}, different models with complex physics implemented obtained quantitatively different results on this particular planet setup.

\begin{figure}[ht]
\includegraphics[width=1.0\columnwidth]{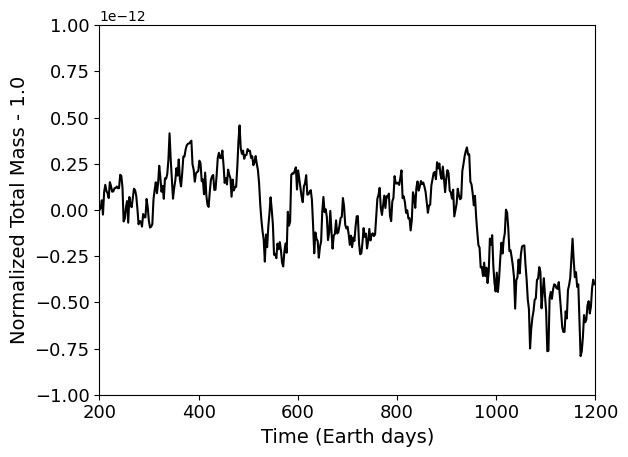}
\caption{Evolution of the total mass error in the ocean tidally locked planet simulation.}
\label{fig:mass_conservation_Exoplanet}
\end{figure}

Our simulation has been integrated for 1200 Earth days, similar to the ocean Earth-like simulation. The time step is 50 seconds, half of the value used for the ocean Earth-like simulation, due to numerical stability issues associated with the different atmospheric dynamics. Again, in this experiment, we discarded the first 200 Earth days as part of the numerical spin-up phase of the simulation.  

Figure \ref{fig:mass_conservation_Exoplanet} shows the total mass error for the ocean tidally locked planet simulation. As for the aqua Earth-like case, the error measured along the simulation is small and not larger than 7.5$\times$10$^{-13}\%$, which is again an example of the good properties of the newly implemented solver conserving atmospheric mass in the 3D simulations.

\begin{figure*}
\begin{centering}
\subfigure[Zonal winds]{
\includegraphics[width=0.9\columnwidth]{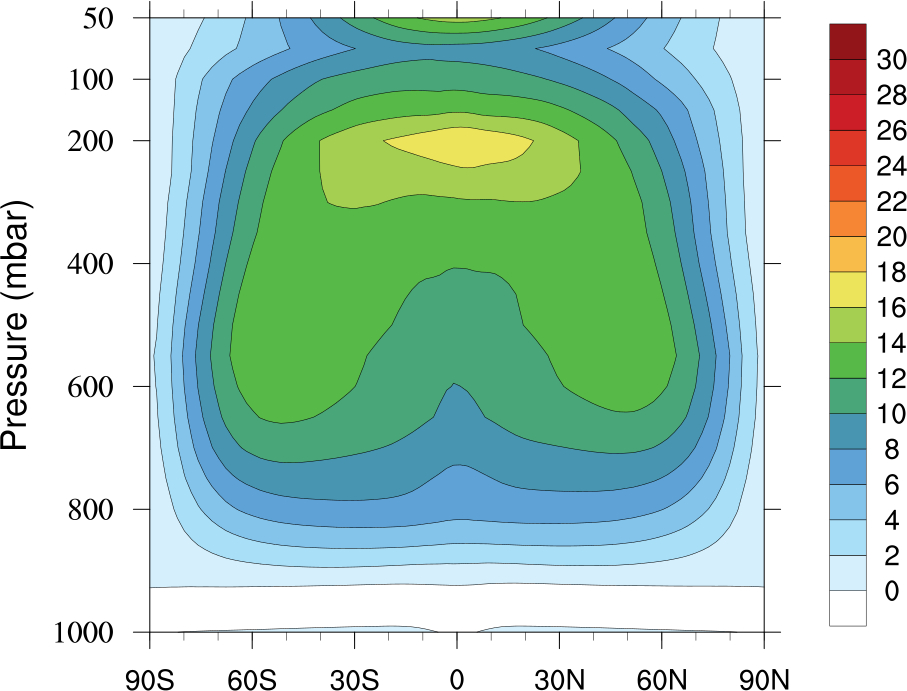}\label{fig:zwinds_temp_exoplaneta}}
\hspace{0.5cm}
\subfigure[Temperature]{
\includegraphics[width=0.9\columnwidth]{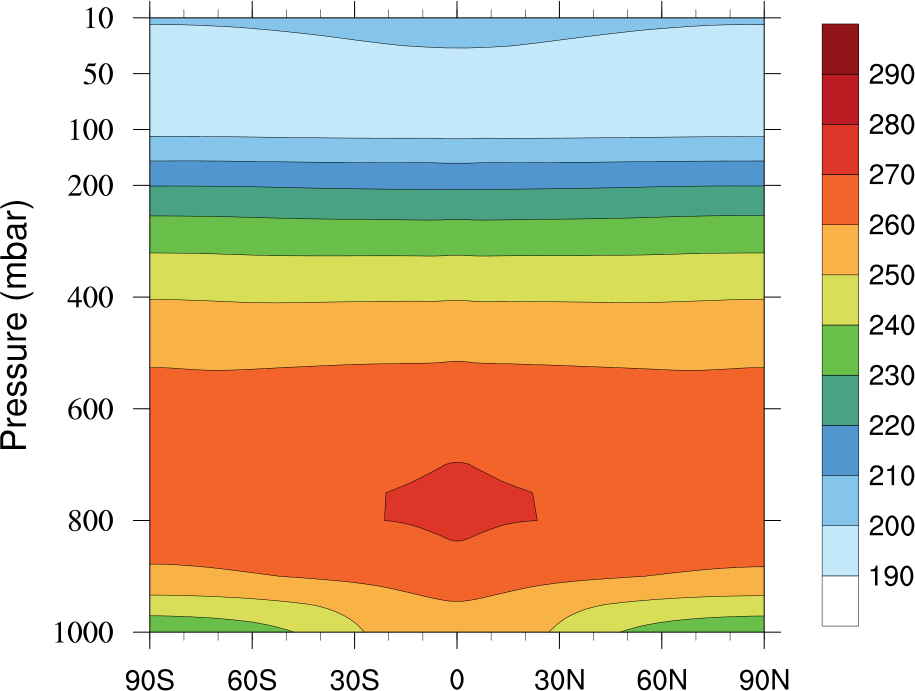}\label{fig:zwinds_temp_exoplanetb}}
\caption{Final time- and longitudinal-averaged zonal winds (m$\,$s$^{-1}$) and temperatures (K) for the ocean tidally locked planet experiment. The values were time-averaged for 1000 Earth days.}
\label{fig:zwinds_temp_exoplanet}
\end{centering}
\end{figure*}

\begin{figure}[ht]
\includegraphics[width=1.0\columnwidth]{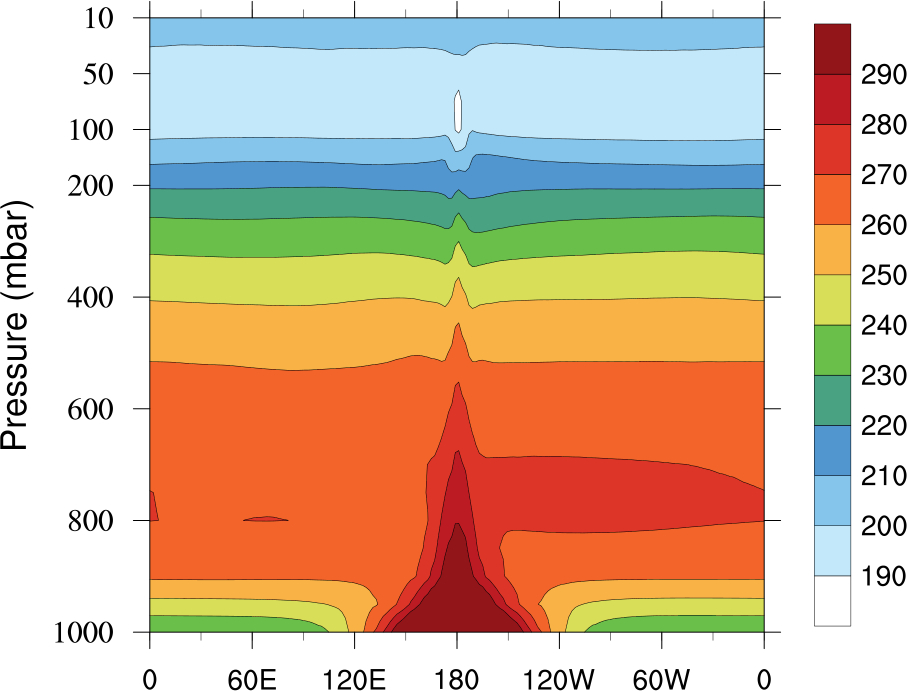}
\caption{Final time- and longitudinal-averaged temperature (K) across the equator for the ocean tidally locked planet simulation. The values were time-averaged for 1000 Earth days.}
\label{fig:t_exoplanet}
\end{figure}

\subsubsection{Zonal winds and temperature}

The tidally locked planet is rotating five times slower than Earth, which strongly impacts atmospheric circulation. In Fig. \ref{fig:zwinds_temp_exoplaneta} we present the averaged zonal winds. The winds obtained are, in general, slower than the winds obtained in the previous experiment. However, as expected from slowly rotating planets, the equatorial winds become stronger (e.g. \citealt{2010Merlis}). The weaker Coriolis force allows more efficient transport of heat from low to high latitudes. The more efficient poleward transport of heat reduces the latitudinal temperature gradients as shown in Fig. \ref{fig:zwinds_temp_exoplanetb} compared to the ocean Earth-like planet case. Another significant difference between the previous experiment and the tidally locked planet case is the vertical gradient of the temperature. In the tidally locked experiment, the atmospheric temperatures reach a maximum at pressures of nearly 800 mbar due to the release of latent heat from water condensation. The latent heat released enhances the upward motion on the dayside of the planet. We note that the sub-stellar point is at 180$^\circ$ longitude. The effect of the latent heat released produces a peaky feature in the temperature structure, as is seen in Fig. \ref{fig:t_exoplanet}. The heat produced on the dayside is transported to the nightside by the prevailing eastward wind, which is driven by the rotation of the planet. The longitudinal heat transport is generally efficient enough to maintain the day-night differences very small, except for pressures higher than roughly 700 mbar. In the lower part of the atmosphere, the simulations showed an asymmetry in the temperature structure between the west and east side of the sub-stellar point. The colder regions of the atmosphere near the surface are due to the very cold temperatures prescribed to reproduce the ocean conditions simulated in \cite{2019Yang}, which reach temperatures below the water freezing point in some regions of the planet.

\subsubsection{Mass stream function}

\begin{figure}[ht]
\includegraphics[width=1.0\columnwidth]{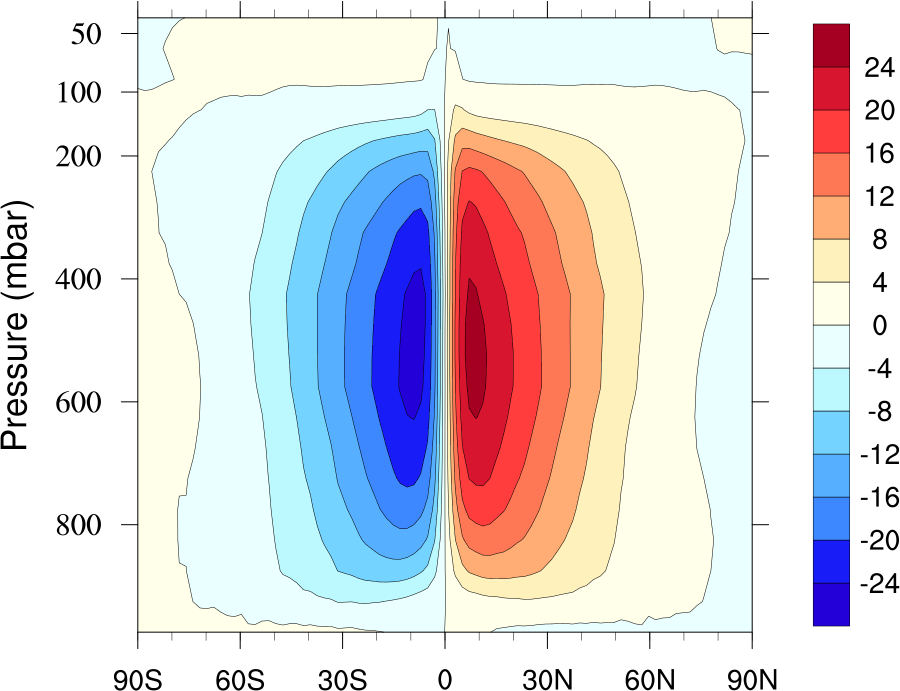}
\caption{Final time- and longitudinal-averaged mass stream function (10$^{10}$kg$\,$s$^{-1}$) for the ocean tidally locked planet simulation. The values were time-averaged for 1000 Earth days.}
\label{fig:psi_exoplanet}
\end{figure}

The averaged mass stream function results are shown in Fig. \ref{fig:psi_exoplanet}. Two large-scale circulation cells form in each hemisphere that extends from the equator to the poles. These cells make the heat transport from low latitudes to the poles more efficient than in the ocean Earth-like experiment. The planet size cells are formed due to the slow rotation of the planet that weakens the impact of the Coriolis force driving the circulation. Large-scale cells such as the ones shown in Fig. \ref{fig:psi_exoplanet} are expected in Venus (e.g. \citealt{2016Mendoncaa}) and Titan (\citealt{2012Lebonnois_Titan}). However, in Titan, the size and position of the large-scale cells evolve in time due to the seasonality effects. 

\subsubsection{Vertical velocity}

\begin{figure}[ht]
\includegraphics[width=1.0\columnwidth]{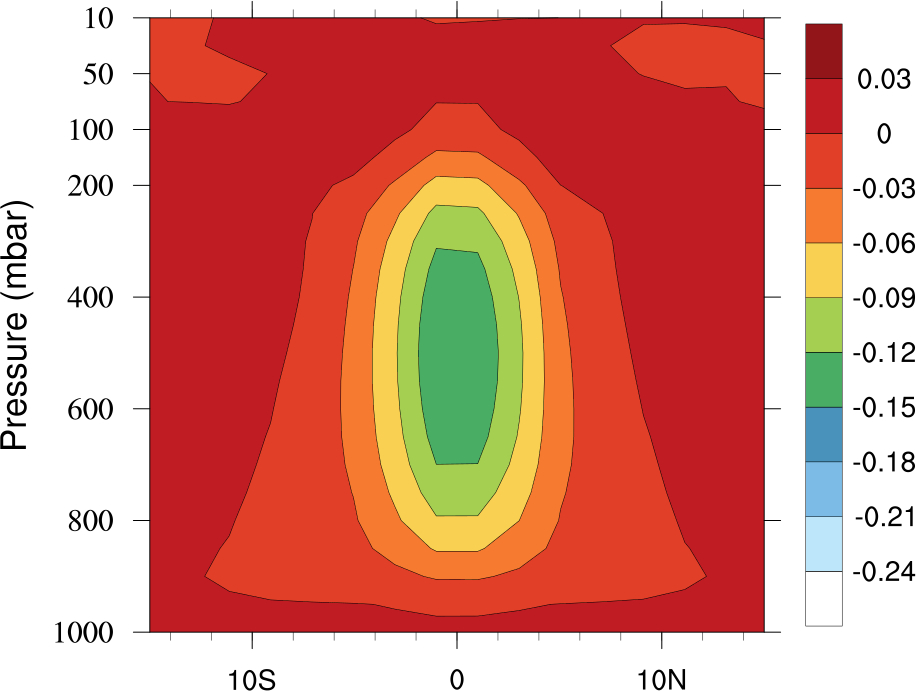}
\caption{Final time- and longitudinal-averaged vertical velocity (Pa$\,$s$^{-1}$) between 15$^\circ\,$S and 15$^\circ\,$N for the ocean tidally locked planet simulation. The values were time-averaged for 1000 Earth days.}
\label{fig:w_exoplanet}
\end{figure}

The vertical component of the winds is shown in Fig. \ref{fig:w_exoplanet}. The distribution of the vertical winds in the equatorial region is similar to the structure obtained in the ocean Earth-like planet experiment. However, the magnitude of the averaged upward winds is stronger in the tidally locked case. The vertical velocities are enhanced in the equatorial region due to the more intense latent heat released from water condensation. The stronger vertical values are located just above the warm region produced by the water condensation and centred at the pressure level 700 mbar.

\subsubsection{Kinetic energy spectrum}

\begin{figure}[ht]
\includegraphics[width=1.0\columnwidth]{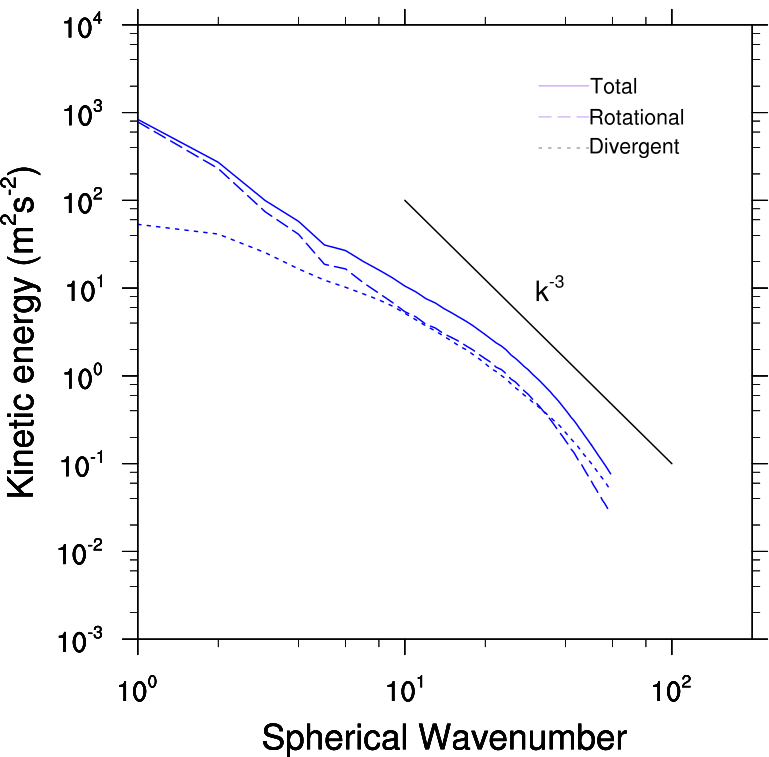}
\caption{Time-averaged kinetic energy spectrum at 0.25 mbar for the ocean tidally locked planet simulation. The dashed and dotted lines are the rotational and divergent components of the kinetic energy, respectively, and the solid line is the total (sum of the rotational and divergent components).}
\label{fig:kspectrum_exoplanet}
\end{figure}

We also computed the kinetic energy spectrum for the tidally locked planet simulation. The kinetic spectrum is shown in Fig. \ref{fig:kspectrum_exoplanet}. Compared with the ocean Earth-like planet case, the presence of the stronger components at shorter wave numbers is clear. As in the Earth-like case, the divergent component of the kinetic energy is much smaller than the rotational component at the smallest wave numbers. However, the two components become comparable at around the spherical wavenumber 10. The characteristics of the temperature forcing (tidally locked configuration) plus the slower rotation of the planet compared to the Earth-like experiment resulted in a stronger generation of divergent kinetic energy. The slope near the truncation limit becomes larger due to the hyper-diffusion and divergent damping processes applied. A buildup of energy indicates that models produce numerical noise that is not being correctly removed. This is not the case in our model, as in the ocean Earth-like simulation, we also do not see a buildup of energy at the truncation limit in this experiment.  The results show that our new methods are robust enough to simulate mass transport in moist atmospheres.

\subsubsection{Water vapour concentration, relative humidity, and precipitation}

\begin{figure*}
\begin{centering}
\subfigure[Water vapour concentration]{
\includegraphics[width=0.9\columnwidth]{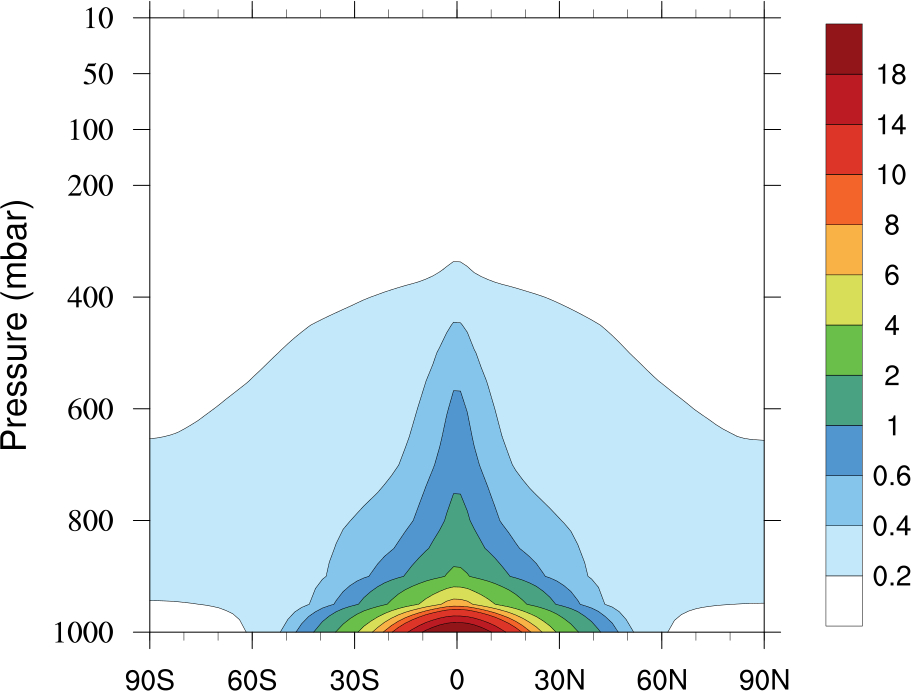}\label{fig:qwater_exoplanet}}
\hspace{0.5cm}
\subfigure[Relative humidity]{
\includegraphics[width=0.9\columnwidth]{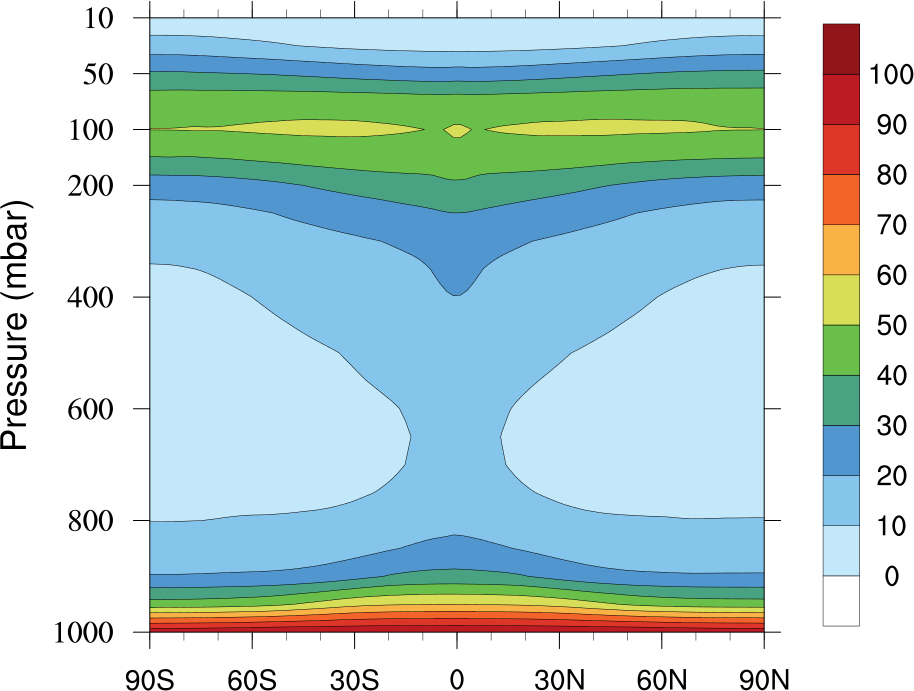}\label{fig:relvh_exoplanet}}
\caption{Final time- and longitudinal-averaged water vapour concentration (kg/kg) and relative humidity ($\%$) for the ocean tidally locked planet simulation. The values were time-averaged for 1000 Earth days.}
\label{fig:q_relh_exoplanet}
\end{centering}
\end{figure*}

The averaged water concentration and relative humidity are shown in Fig. \ref{fig:q_relh_exoplanet}. The largest concentration of water vapour is located at low latitudes in the lower atmosphere. As expected from the temperature distribution, we also obtain an elongated vertical distribution at the equator due to the enhanced acceleration of the vertical winds by latent heat released from water condensation. Above the pressure level of 400 mbar, the concentrations of water vapour become very small. Figure \ref{fig:q_relh_exoplanet} shows low levels of relative humidity in the lower atmosphere, which is caused by precipitation. Higher relative humidity values are located near the surface and in a pressure region between 50 and 200 mbar. The relatively high values of relative humidity at low pressures seem to indicate that the atmosphere is not efficient at removing water vapour at these altitudes since this region is not cold enough to condense water and vertical mixing is not efficient.

\begin{figure}[ht]
\includegraphics[width=1.0\columnwidth]{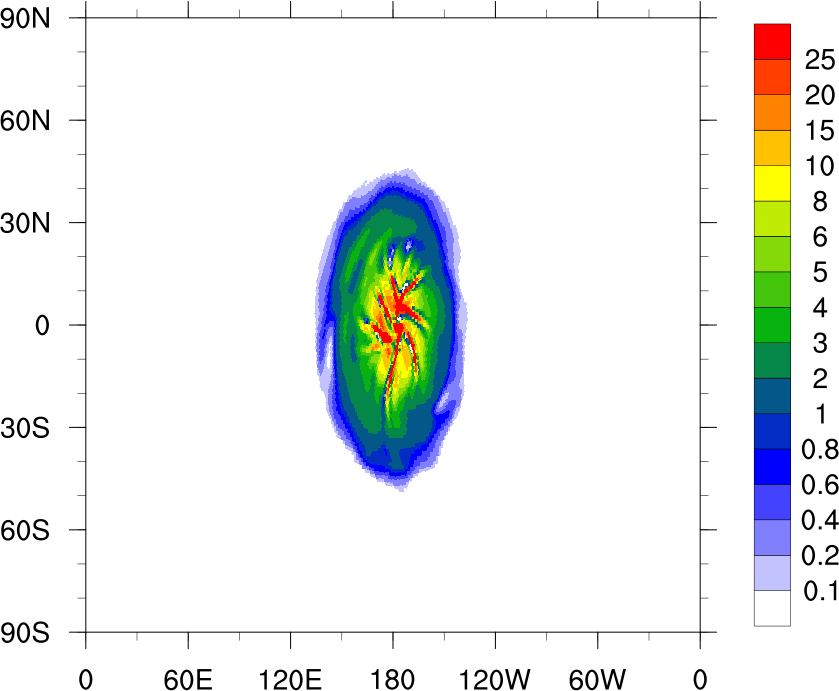}
\caption{Instantaneous latitude-longitude map of surface precipitation (mm$\,$h$^{-1}$) for the ocean tidally locked planet simulation.}
\label{fig:percipitation_exoplanet}
\end{figure}

In Fig. \ref{fig:percipitation_exoplanet} we show the surface precipitation. Our results show a permanent precipitation region on the dayside of the planet. The complete water cycle is only possible on the dayside, where a large amount of water is evaporated from the hot dayside surface regions, transported upwards and precipitates. This cycle does not extend to the nightside. This configuration is a characteristic of the parameter space explored, and it does not mean that permanent precipitation on terrestrial tidally locked planets will only happen on the dayside. A more thorough exploration of the parameter space (e.g. rotation rate and surface pressure) needs to be done in the future. The precipitation map does not register any possible indication of numerical error, which would be noticed by a random distribution of maxima in the precipitation. The precipitation structure forms a round shape structure with spiral arms, which resembles of tropical cyclones on Earth. However, the formation of the large-scale precipitation structure on the dayside of the tidally locked planet is different from Earth tropical cyclones, which are mainly driven by rotational effects. The tidally locked planet large-scale precipitation region is formed mostly due to divergent processes triggered by the release of latent heat from the water condensation and the permanent dayside configuration.

\section{Conclusions}
\label{sec:concl}
The formulation presented in this work is part of the ambitious development of the PCM \texttt{OASIS} (\citealt{2020Mendonca}) that has been built from scratch to explore a large diversity of planets. \texttt{OASIS} includes various physical and chemical schemes in a modular approach that allows planetary climate problems to be explored at different levels of complexity. Every module implemented in the \texttt{OASIS} platform is carefully assessed in terms of performance, accuracy, physical consistency, and flexibility to explore an extensive range of planetary conditions. A complete design of a condensible cycle in 3D models needs to include methods to represent, for example, mass transport, cloud formation, cloud radiative feedbacks, and moist atmospheric thermodynamics. In this work, we focus on the implementation of the first step towards the complete implementation of a robust condensible cycle in 3D simulations: the mass transport scheme.

We have implemented an upwind-biased scheme on a piece linear approximation with a flux limiter from \cite{2007Miura} to represent the mass transport in the 3D simulations. The new numerical method has been successfully tested in problems with different complexity. We first explored the new solver on a 2D model where we transport a bell-shaped cosine distribution across the planet. Compared to the old central finite volume scheme, the results of the new solver show a significant improvement in terms of, for example, mass conservation, accuracy, shape-preserving properties, and avoiding negative solutions. The new solver has a second-order accuracy and avoids using any fixer, such as hyper-diffusion applied to the mass variable, which improves the physical consistency in the model. We compared the 2D results for different horizontal resolutions and showed that for 1-degree-resolution simulations the error is less than 1$\%$ for one complete revolution of the peak. The error needs to be kept as small as possible to avoid any unphysical feedback processes in the simulation. We recommend using at least 1-degree resolution for moist atmospheric simulations to avoid impactful errors and a low degree of implicit numerical diffusion.

Another goal of this work was to test our new formulation on 3D simulations that include a simple physical representation of the moist processes in the atmosphere. To accomplish this goal, we benchmarked our model against a test developed in the Earth climate community and propose a new tidally locked terrestrial planet benchmark test. 

Our new model was capable of passing the ocean Earth-like benchmark test successfully. The results obtained with our new \texttt{OASIS} model are quantitatively similar to the results presented in \cite{2016Thatcher}, who designed the test. The simulations produce robust representations of atmospheric mass transport and consistent water vapour distributions with more complex Earth climate models that include sophisticated schemes to represent cloud formation. It is well known that moist physics tendencies in 3D numerical simulation models can trigger undesirable large-scale gravity waves (e.g. \citealt{2016Thatcher}), which were not detected in any of our simulations. Our results also showed that our new formulations do excellent work conserving the total atmospheric mass (error is less than $5\times10^{-13}\%$).

We offer the exoplanet community a robust new benchmark test for 3D exoplanet models to simulate ocean tidally locked planets. The new test explores the challenges of simulating a moist atmosphere of a tidally locked planet. The bulk parameters of our tidally locked planet are based on the simulations studied in \cite{2019Yang}, who explore a tidally locked planet orbiting an M star. Our results show a planet where water evaporates from the permanent dayside ocean and precipitates on the same side. Condensation occurs in the lower atmosphere and has a stronger impact on the temperature and wind field than in the ocean Earth-like case. In addition, the slower rotation of the planet drives the atmospheric circulation to produce stronger winds at the equatorial region, which creates permanent strong easterly winds that have a substantial impact on the transport of water vapour from the dayside to the nightside. Our simulations showed a good performance of the new solver in terms of accuracy and does not produce any signs of undesirable numerical high-frequency waves (e.g. numerical noise). Furthermore, the benchmark is easy to set up. We encourage other groups to carry out our proposed test as a first step in evaluating the methods used to explore the exotic planet environments present in ocean tidally locked terrestrial planets. 

This work is the first part of a series of manuscripts that will focus on the full implementation of moist processes in 3D planetary models. We will present the complete formalism implemented in the \texttt{OASIS} model and propose benchmark tests with different levels of complexity to evaluate all the various steps of the numerical implementation. 

\begin{acknowledgements} 
J.M.M. acknowledges financial support from the PRODEX project number 4000127377. \texttt{OASIS} was run on the HPC cluster at the Technical University of Denmark (\citealt{DTU_DCC_resource}). The figures in this study have been done with the NCAR Command Language (Version 6.6.2) [Software]. (2019). Boulder, Colorado: UCAR/NCAR/CISL/TDD. http://dx.doi.org/10.5065/D6WD3XH5
\end{acknowledgements} 

\bibliographystyle{aa} % style aa.bst

\end{document}